\documentclass[11pt, a4paper, oneside]{article}

\usepackage{amssymb}
\usepackage{amsmath}

\usepackage{graphicx} 
\usepackage{amsmath,amsthm, amssymb, latexsym}
\usepackage{multirow}
\usepackage{todonotes}
\usepackage{url}
\usepackage{hyperref} 
\usepackage{tikz}
\usetikzlibrary{arrows.meta}
\usepackage[table]{xcolor} 
\definecolor{darkgreen}{HTML}{006400}
\usepackage{rotating}
\usepackage{subfig}
\usepackage[utf8]{inputenc}   
\usepackage[T1]{fontenc} 
\usepackage{booktabs} 

\begin{document}



\title{EMU circulation planning for Silesian Railways: case study and a quantum approach}


\author{Ewa K\k{e}dziera\thanks{ekedziera@iitis.pl, Institute of Theoretical and Applied Informatics, Polish Academy of Sciences, Gliwice}, \
Wojciech Gamon\thanks{ORCID: 0000-0002-6015-2597, Silesian University of Technology, Department of Railway Transport, Faculty of Transport and Aviation Engineering, Gliwice, Poland}, \
 M\'aty\'as Koniorczyk\thanks{ORCID: 0000-0002-2710-493X, HUN-REN Wigner Research Centre for Physics, Department of Quantum Optics and Quantum Information, Budapest, Hungary}, \
 Zakaria Mzaouali\thanks{ORCID: 0000-0003-3948-1318, Eberhard Karls Universit\"at T\"ubingen, Institut für Theoretische Physik, Tübingen, Germany; Forschungszentrum Jülich, Jülich Supercomputing Center, Jülich, Germany}, \\
 Andrea Galad\'ikov\'a\thanks{University of \v{Z}ilina, Department of Mathematical Methods and Operations Research, Faculty of Management Science and Informatics, \v{Z}ilina, Slovakia}, \
 Krzysztof Domino\thanks{kdomino@iitis.pl, ORCID: 0000-0001-7386-5441, Institute of Theoretical and Applied Informatics, Polish Academy of Sciences Ba{\l}tycka~5, 44-100 Gliwice}}

\maketitle
\begin{abstract}
We study daily rolling stock circulation planning for electric multiple units (EMUs) on a regional passenger network, focusing on services where identical EMUs may be coupled in pairs on selected routes. Motivated by the operational needs of the regional operator Silesian Railways in Poland, we formulate an acyclic mixed‑integer linear program on a one‑day horizon that incorporates depot balance constraints, demand‑driven seat and bicycle capacity limits, and simple crew availability constraints.  Using a graph/hyper-graph representation of train movements, we first solve an ILP formulation. We then derive a Quadratic Unconstrained Binary Optimization (QUBO) reformulation and evaluate its solution by quantum annealing on D‑Wave Advantage systems and by the classical quantum‑inspired VeloxQ solver. In computational experiments on real‑world instances from the Silesian network, with up to 404 train trips and 11 EMU types the ILP approach yields high‑quality daily circulation plans within at most about 40 minutes. The quantum and quantum‑inspired solvers are restricted to substantially smaller sub‑instances due to the large number of terms in the QUBO, and embedding limitations in the case of quantum hardware. These results quantify the present frontier of QUBO‑based methods for rolling stock circulation. They can be helpful in designing a hybrid classical-quantum approach.
\end{abstract}

\begin{figure}[h!]
    \centering
    \includegraphics[width=\linewidth]{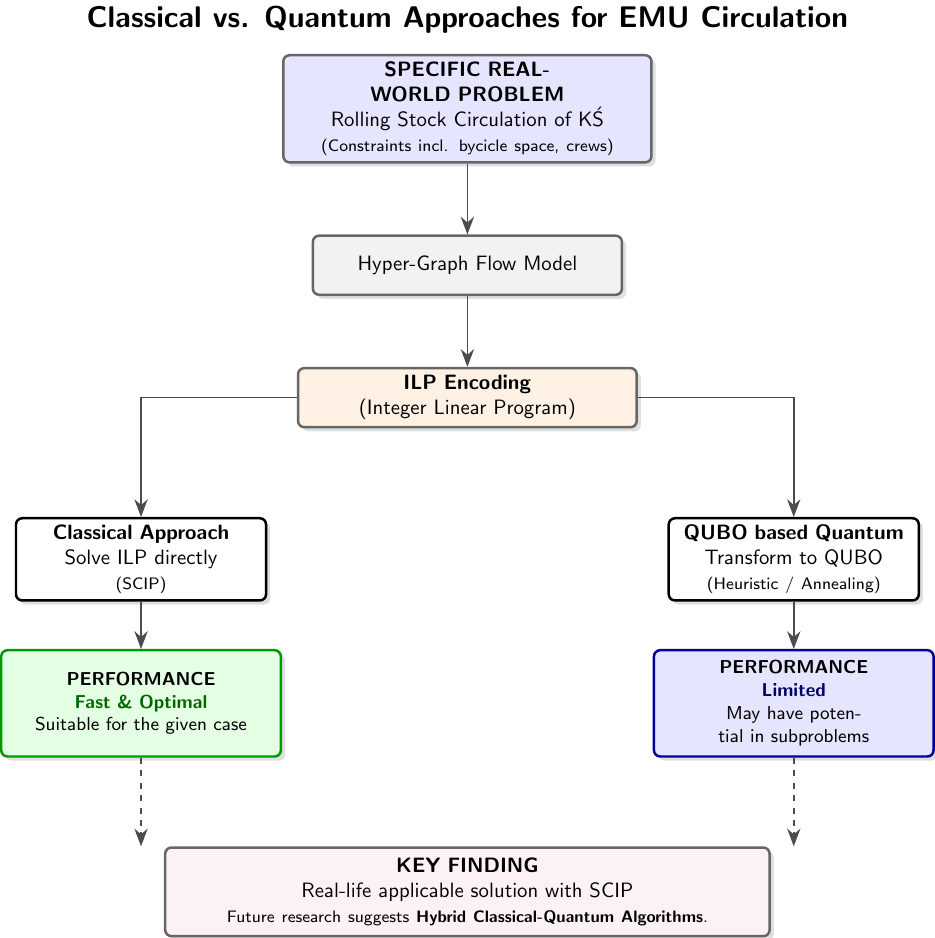}
    \label{fig_GA}
\end{figure}

\section*{Keywords}
Rolling stock circulation planning; stakeholder-driven approach; passenger and bicycle capacity; quantum annealing; VeloxQ QUBO solver


\section{Introduction}

Rolling stock circulation planning is an essential component of passenger railway operations management~\cite{Caprara2007}. Given a passenger timetable and the characteristics of the available rolling stock, the goal is to find an optimal assignment of vehicles to scheduled trips such that all operational, technological, and capacity requirements are satisfied. Depending on the planning context, the optimization criteria may include, among others, minimizing operating costs, reducing empty runs, ensuring sufficient passenger capacity, or maintaining adequate rolling stock reserves. The problem has many variants and has been extensively studied in the literature, as discussed in Section~\ref{subsect:literature}.

In this paper, we address a rolling stock circulation problem motivated by the operational requirements of a regional passenger rail operator, Silesian Railways (Polish: Koleje Śląskie). The operator's services are performed by electric multiple units (EMUs), which usually return to their depots on a daily basis. Depots may hold spare units that serve as an operational reserve and play an important role in short-term planning. In addition to rolling stock availability, the generated plan must remain compatible with crew-related operational requirements, including the availability and suitability of train drivers for particular services.

The aim of this work is to develop a flexible optimization model for rolling stock planning. The model is designed to support alternative optimization objectives and to balance two main goals: maintaining rolling stock reserves at depots and minimizing the operational costs associated with rolling stock usage. In addition, motivated by passenger feedback collected by the operator, we include bicycle capacity requirements as an explicit constraint in the model. To the best of our knowledge, this particular combination of operational features has not been addressed in the existing literature on passenger railway rolling stock circulation optimization.

From a modelling perspective, rolling stock circulation problems are commonly formulated as variants of set-flow models; when flexible train compositions are involved, these naturally extend to hypergraph formulations, as adopted in this paper. Such formulations can become computationally demanding as the size of the planning instance increases. Nevertheless, for the instances considered in this study, we obtain practically useful solutions using a state-of-the-art solver for constrained integer programming. This classical approach serves as the main reference point for evaluating alternative solution methods.

As an experimental direction, we also reformulate the proposed model as a Quadratic Unconstrained Binary Optimization (QUBO) problem. The reformulation uses penalty terms and slack variables, which are standard tools for transforming constrained integer programs into QUBO form~\cite{glover2018tutorial}. The resulting QUBO instances are solved using both a hardware quantum annealer and a solver based on advanced physics-inspired heuristics. This comparison allows us to assess the current practical frontier of QUBO-based quantum and quantum-inspired solvers, applied to hypergraph-based rolling stock circulation planning, relative to the classical integer programming approach.

The problem considered in this paper arises from the current planning practice of the railway operator. Rolling stock schedules have traditionally been prepared by the operator's staff with limited IT support. Using their knowledge of the timetable, available EMUs and their technical and operational characteristics, depot-based driver pools, and expected passenger and bicycle demand, planners prepare detailed planning tables. The resulting planning tables assign individual EMUs, or pairs of EMUs in predefined cases, to scheduled trips operated on specified routes and at specified times, while respecting internal and external operational constraints.

The proposed model is intended to support this planning process by generating a daily rolling stock circulation plan. We focus on rolling stock planning over a one-day operational horizon and consider instances of different sizes, with up to \(404\) scheduled trips. Since trips are ordered in time within the planning horizon, the resulting circulation structure is acyclic. The model can also be used to generate alternative plans under modified operational conditions, such as increased passenger or bicycle demand, rolling stock unavailability, or other disturbances, hence it is also suitable for reactive planning to some extent.

The generated plan must satisfy several operational requirements that reflect the operator's planning practice. All scheduled trips must be covered, and the continuity of rolling-stock circulation must be maintained throughout the day. The plan must also satisfy prescribed bounds on the number of EMUs of each type that start and end the day at each relevant location. In addition, the assigned rolling stock must provide sufficient capacity to meet both passenger and bicycle demand. Finally, the plan must respect aggregate crew-related requirements, in particular limits on the number of suitable drivers available at a given time.

The model also allows selected trips to be operated either by a single EMU or by two coupled EMUs. Coupling is permitted only under predefined restrictions: the coupled EMUs must be of the same type, and coupling is allowed only on trips specified by the operator.

Operating costs consist of fixed costs and distance-dependent variable costs. This cost term is combined with the number of EMUs dispatched, controlled by a weighting parameter, which allows the operator to analyse different trade-offs between rolling stock reserve and operating cost.

Crew requirements are included only at an aggregate level. The model does not solve the crew rostering or crew scheduling problem. Instead, it ensures that the vehicle assignment is compatible with the available number of suitable drivers, so that a detailed crew roster is likely to remain feasible in a subsequent planning step.

\subsection{Literature review}
\label{subsect:literature}

The optimization of railway rolling stock circulation is one of the fundamental tasks in passenger railway transport planning. Due to the strong interdependence between the timetable structure, rolling stock availability, and operational constraints, a wide range of different approaches to solving this problem has emerged, particularly within mathematical programming. In what follows we review the part of this extensive literature which is the most pertinent to our research.

A traditional starting point for modelling rolling stock circulation
is represented by constrained network flow formulations \cite{Lamorgese2018}, either in arc-based or path-based form, which are often solved using advanced decomposition techniques or column generation.  
Fioole et al. \cite{fioole2006rolling} are among the first works to address rolling stock circulation in detail.
Theirs can be considered as a structural ancestor of the present contribution.
It also addresses a daily planning horizon, and presents a mixed-integer linear model with weighted objective balancing operating cost and service quality.
Importantly their model incorporates the composition of electric multiple units: coupling and decoupling operations are represented, while respecting temporal and station-related constraints.
An important element of this work is also the introduction of a seat-shortage indicator, which links rolling stock circulation planning with the quality of service provided to passengers. This formulation is therefore close to our approach, especially because it combines technical decisions on unit circulation with capacity requirements.
However, the coupling of units we have to model is different: our couplings are restricted to predefined EMUs and trips.
Their schedule is cyclic while ours is acyclic, and they do not consider certain constraints such as, depot balance per type or crew constraints.

An important topic in the literature is the flexible composition of electric multiple units.
From an operational perspective, this option is significant because it allows train capacity to be adjusted to expected demand.
At the same time, however, it considerably complicates the optimization problem, since the model must decide not only which vehicle will serve a particular train service, but also whether the service will be operated by one unit or by several coupled units.

When multiple-unit trainsets or flexible train compositions are included in the model, a simple network representation may no longer be sufficient.
In such cases, it is necessary to model not only the transition of a single vehicle between two train services, but also combined decisions involving several units at once.
This naturally leads to hypergraph formulations.
Borndörfer et al.~\cite{Borndorfer2021} propose a hypergraph-based approach in which hyperarcs represent feasible compositions of multiple train units.
The model makes it possible to account for constraints related to platform length, trainset orientation, and the operational feasibility of multiple-unit compositions. The solution is based on hierarchical column generation, which gradually works with different levels of detail.
Comparing with the present contribution, here we also adopt a hyperarc-based representation, but we do not adopt hierarchical column generation: owing to the smaller scale of our problem, our ILP can be solved directly.
While Borndörfer et al.~\cite{Borndorfer2021} also consider multiple vehicle types and depots, they have no demand-driven capacity constraints, and their couplings are not restricted to a trip subset.
In addition, the present paper is the first to combine the hyperarc representation with QUBO.

We remark here that the application of quantum methods in rolling stock circulation planning is limited.
To our knowledge, the closest contribution is the article by Grozea et al.~\cite{grozea2021optimising}, addressing rolling stock circulation planning on the German Railways network.
That problem, however, differs significantly from the one considered in this paper with respect to input data, output requirements, and constraints.
Their formulation does not include passenger capacity constraints or demand satisfaction constraints, assumes a homogeneous train fleet, and assigns trips to single train units; thus, their comparison of classical and quantum approaches does not cover the type of models studied here.

Concerning passenger capacity requirements --- an important aspect of rolling stock circulation planning --- in the early work of Peeters et al.~\cite{PEETERS2008}, demand is incorporated into the design of a branch-and-price algorithm in such a way that it influences the generation of feasible columns itself.
Capacity requirements are therefore not merely checked after a solution has been created, but are entered directly into the process of generating candidate circulations.
This approach is methodologically strong for large-scale problems, since it is not necessary to explicitly generate all possible circulations in advance.
At the same time, however, it requires a specialized solution framework.

Our contribution remains within the realm of hierarchical planning: we treat the timetable as a fixed input to rolling stock circulation planning.
The contributions summarized in what follows deal with integrated timetable-rolling stock circulation modeling; although they differ from our fixed-timetable approach, they provide relevant context.

The relationship between passenger demand and rolling stock circulation is even more pronounced in models that use a multi-stage or iterative structure.
Wang et al.~\cite{WANG2018} propose and compare iterative nonlinear and MILP-based solution approaches for the joint planning of train scheduling and rolling stock circulation under time-varying sectional demand; the MILP reformulation yields the best results.
The result is a gradual alignment of capacity decisions and vehicle circulations.
This work shows that passenger demand should not be perceived merely as a static constraint, but as a factor that significantly influences the structure of the solution itself.
Meanwhile it should be noted that this contribution addresses a different railway context: a metro line.
There are no constraints on depot balance per type or crews there, and the trains have fixed composition.

Another important line of research is represented by exact decomposition approaches, especially column generation and branch-and-price. Nishi et al.~\cite{NISHI2017} use a combination of column generation and Lagrangian relaxation to solve a rotation problem.
Their model works with decisions concerning transitions between train services within rotation cycles, embedding periodic inspection constraints that require each trainset to be serviced at least once every given number of days.
Pan et al.~\cite{PAN2023} extend this type of approach to problems with flexible train compositions and use a heuristic based on column generation and a diving strategy.
These works show that column generation is particularly suitable when the solution can be composed of larger structures, such as complete circulations or rotation blocks. Its disadvantage, however, is the need for a specialized algorithmic framework, which is usually more difficult to implement than a direct integer formulation.

The authors in \cite{ZHOU2022} propose a MILP model that simultaneously optimizes the railway timetable and rolling stock circulation plan, while allowing changes in trainset composition at terminal stations; the model is solved using Variable Neighbourhood Search rather than column generation.
This work, applied to a metro line with tidal demand, is another instance of the broader research direction that couples rolling stock composition decisions with timetabling --- in contrast to our fixed-timetable assumption.

Pan et al.~\cite{PAN2024} advance this topic further by addressing the planning of flexible train compositions under stochastic passenger demand.
Demand uncertainty is practically important in railway transport, since the actual number of passengers may differ from expected values.
The authors use Benders' decomposition, which makes it possible to divide a large-scale mixed-integer problem into more manageable parts.
Their work is important because it combines flexible train composition, demand-oriented planning, and uncertainty.
At the same time, it shows that more realistic demand modelling often requires advanced decomposition methods and longer computational times.
Their model applies to a metro line where stochastic passenger demand can bear significant relevance.
In our case the demands can be predicted to a sufficient extent, not necessitating the overhead implied by stochastic methods. 

The works of Pan et al.~\cite{PAN2023, PAN2024} are the closest ILP-based analogues in terms of problem structure (flexible composition and demand), but the joint-timetabling assumption makes them a different problem.
Our paper's fixed-timetable assumption is a conscious pragmatic choice that keeps the model tractable for the regional operator — one research gap our paper addresses.

Heuristic and metaheuristic methods are also used to solve computationally demanding integrated problems.
Perhaps the most recent contribution that considers flexible compositions is by Bao et al.~\cite{BAO2025} who use simulated annealing to address a real-world metro case study. 
They also have demand-oriented capacity constraints, but no constraints on crews.
Like the quantum and quantum-inspired solvers we employ for QUBO instances, simulated annealing methods do not provide proof of optimality, but they can be practically useful in situations where it is important to obtain a good solution within a reasonable time.
Their results often depend on algorithm parameter settings, the quality of the proposed neighborhoods, and the way the solution space is searched.

Exact graph-based methods provide a complementary perspective.
Gao et al.~\cite{GAO2022} develop a branch-and-price algorithm based on a rotation network of event nodes for weekly planning in high-speed railway transport.
In their formulation, nodes represent operational events, such as train starts and ends, maintenance activities, or idle periods, while arcs encode possible transitions between these events and capture the flow of train units between different operational states.
The resulting graph is then encoded as an ILP model, in which binary decision variables select arcs forming feasible weekly rotations.
The arc/rotation-network structure is related to our graph/hyper-arc model, but their weekly horizon introduces cyclic rotation structure that requires branch-and-price; our acyclic daily model avoids this.
They do not have demand-driven capacities.

Multi-stage approaches also frequently appear in connection with operational robustness and rescheduling.
Haahr et al.~\cite{HAAHR2016} address daily rolling stock scheduling and short-term rescheduling under disruptions using MILP and column-and-row generation.
Their approach minimizes costs related to composition assignments, composition transitions, and end-of-day balance penalties, and is able to handle large-scale instances, although it does not include flexible trainset composition or demand-based capacity constraints.
In our case, however, rescheduling can be performed solving the ILP or QUBO directly, without column generation.

A separate group of studies addresses integrated railway planning problems using multidimensional network representations.
In \cite{CHAI2025}, the need for joint modelling of periodic and non-periodic timetables together with rolling stock circulation is emphasized.
The authors use a three-dimensional time-space-state network, which captures not only time and location, but also the state of a vehicle or trainset.
Due to the high dimensionality of the resulting model, the solution is based on Lagrangian relaxation.
Niu et al.~\cite{NIU2025} use a similar multidimensional network structure and apply ADMM decomposition to coordinate the individual subproblems.
These works are significant mainly from a methodological point of view, as they show how large-scale and strongly interconnected planning problems can be handled.
For practical daily planning by a regional operator, however, such models may be unnecessarily complex, especially when the timetable itself is fixed.

In summary, the literature on rolling stock circulation offers a wide range of modelling paradigms, including network-flow formulations, mixed-integer models with flexible train compositions, decomposition-based approaches, and large-scale exact and metaheuristic methods.
While many studies address capacity constraints, passenger demand, or operational feasibility, only a limited subset simultaneously considers flexible EMU compositions together with demand-driven capacity requirements in a way that remains computationally tractable for real-world instances.
Bike capacity constraints are, to the best of our knowledge, a new element here: they were not considered in the literature before.
These observations naturally lead to several research gaps, which we discuss in the following section.

\subsection{Research gaps and contribution}

We recognize the following research gaps. From the railway application perspective, there is no model available in the literature with the exact inputs and outputs required by the industrial railway operator.
While our model belongs to the general family of rolling stock circulation models, it incorporates several operator-specific requirements.
In particular, no existing model simultaneously incorporates pairwise same-type EMU coupling restricted to a predefined trip subset, explicit bicycle capacity constraints, depot balance per vehicle type, and aggregate crew-availability constraints within a single directly-solvable ILP formulation.
It is restricted to rolling stock circulation planning for a given timetable.
An important practical feature of the formulation is that it remains tractable for the considered real-world instances without resorting to column generation.

As for quantum computing for railway applications, this study explores a practice-driven rolling stock circulation problem.
Despite this potential, the literature on quantum computing and quantum-inspired approaches for rolling stock circulation planning remains limited to the aforementioned contribution of~\cite{grozea2021optimising}, which addresses a homogeneous fleet without passenger capacity constraints, using a proprietary hybrid solver.

As quantum devices are small and prone to error, a hybrid quantum-classical approach~\cite{DwaveHSS} has been applied therein and to a somewhat similar railway rescheduling problem~\cite{koniorczyk2025solving}.
However, such hybrid solvers are proprietary, and it is difficult to assess the potential impact of the quantum or quantum-inspired part of the algorithm on the overall optimization performance.
This paper therefore compares a pragmatic classical ILP approach with QUBO-based approaches solved using both quantum annealing hardware and a physics-based quantum-inspired heuristic.

Our contributions are: 
\begin{itemize}
    \item practical ILP/hyper‑graph model with EMU coupling on predefined trip subsets and demand‑driven capacity (passenger and bicycle),
    \item QUBO mapping and scaling analysis specific to rolling stock,
    \item empirical comparison ILP vs D‑Wave quantum hardware vs VeloxQ quantum-inspired heuristics~\cite{tuziemski2025veloxq, tuziemski2025recentquantumruntimedisadvantages, hanussek2025solvingquantuminspireddynamicsquantum, Robertson_2025, pawlowski2025closingquantumclassicalscalinggap} on real data, and
    \item implications for future hybrid architectures.
\end{itemize}
Table~\ref{tab::literature_ref} provides a brief summary of the similarities and differences of the present contribution to the ones in the literature most similar to it.

\begin{table}[ht!]
\centering
\caption{Comparison of representative rolling stock circulation models (key features).  \textbf{Legend:}
H = planning horizon (Day = daily, Week = weekly);
C = EMU coupling/splitting;
D = demand-driven capacity constraints;
M = stochastic or multi-stage structure;
$\checkmark$ = fully modeled,
$\checkmark^{*}$ = modeled with predefined or restricted coupling locations,
$\circ$ = limited or scenario-based,
-- = not considered;
ILP = integer linear programming,
MILP = mixed integer linear programming,
INP = iterative nonlinear programming,
BD = Benders decomposition,
CG = column generation,
HG = hyper-graph.}
\renewcommand{\arraystretch}{1.15}
\setlength{\tabcolsep}{4pt}
\begin{tabular}{lccccc}
\hline
\textbf{Reference}
& \textbf{H}
& \textbf{C}
& \textbf{D}
& \textbf{M}
& \textbf{Method} \\
\hline
Fioole et al.~\cite{fioole2006rolling}
& Day & \checkmark & \checkmark & -- & ILP \\

Wang et al.~\cite{WANG2018}
& Day & -- & \checkmark & \checkmark & INP/MILP \\

Pan et al.~\cite{PAN2024}
& Day & \checkmark & \checkmark & \checkmark & MILP+BD \\

Borndörfer et al.~\cite{Borndorfer2021}
& Week & \checkmark & -- & -- & HG+CG \\

Grozea et al.~\cite{grozea2021optimising}
& Day & -- & -- & -- &  QUBO \\

\textbf{This work}
& Day & \checkmark$^{*}$ & \checkmark & $\circ$ & HG+ILP/QUBO \\
\hline
\end{tabular}
\label{tab::literature_ref}
\end{table}
\noindent

This paper is organized as follows.
Section~\ref{sec::model} introduces the proposed hypergraph-based formulation, its ILP encoding, and the corresponding QUBO reformulation.
Section~\ref{sec::solution} describes the computational setup and reports results for real-world instances, comparing the classical ILP approach with quantum annealing and the quantum-inspired solvers.
Finally, Section~\ref{sec::discussion} discusses the practical implications of the results, the current limitations of QUBO-based approaches, and directions for future hybrid classical--quantum methods.

\section{Model}\label{sec::model}

In this section, we present a mathematical model for the railway problem under study, namely EMU circulation planning within the operational time horizon, using an acyclic time-expanded representation. Technically, we encode the problem using a graph-based approach inspired by the \emph{Graph Flow} method discussed in chapter~10 of ~\cite{Lamorgese2018}. In our model, we use the graph/hyper-graph approach, representing trips as nodes, EMU transfers between trips as arcs, and transfers of coupled EMUs as hyper-arcs.

\subsection{Timetable and trips}

 Our problem is defined by a timetable $T$, composed of train trips $v~\in~T$. By a trip, we mean a train journey from its origin station to its destination station; hence, we use the terms train and trip interchangeably throughout the paper. Each trip has a departure time $t_{\text{start}}(v)$ and arrival time $t_{\text{end}}(v)$. For each trip, the model considers only its departure from the origin station and arrival at the destination station; intermediate stops are ignored. Such an approach limits the model’s size and increases the size of tractable problems.

 To cover the timetable, we consider EMUs of various types  and depots. In accordance with the requirements of our industrial partner, EMU coupling is permitted only subject to the following operational restrictions. We consider only couplings of EMUs of the same type, with each coupled composition consisting of at most two EMUs. Couplings are allowed only for a selected subset $T'' \subset T$ of trips, predefined according to the partner’s requirements. A trip $v'' \in T''$ may be served either by a single EMU or by a coupled pair of EMUs. Conversely, let $T'$ denote the set of trips for which coupling is not permitted; each trip $v' \in T'$ must be served by a single EMU. Thus, $T = T' \cup T''$ and $T' \cap T'' = \emptyset$.

\subsection{Hyper-graph representation}

The problem is mapped to a (hyper-)graph composed of nodes \(v \in V\), representing trips from the timetable $T$ and auxiliary depot service trips, and arcs (hyper-arcs) representing transitions of EMUs between trips. Here, the node $v_d$ represents an EMU departing from depot $d$, while the node $v'_d$ represents an EMU entering depot $d$.

In each node, we store train-specific data, namely passenger and bicycle demand, allowed rolling stock compositions, whether the trip is mandatory or optional, and the trip length (for cost calculations). 

An arc \(h_{v,v',r}\) represents a feasible transition of an EMU of type $r$ from trip $v$ to trip $v'$, where $r \in R$ and $R$ denotes the set of EMU types. Hyper-arcs represent analogous feasible transitions for coupled EMUs. We assign to each (hyper-)arc fixed and variable costs, as well as the seating and bicycle capacities of the trips it points to. 

As represented in Fig.~\ref{fig:hyperarcs}(a)-(c), three types of hyper-arcs are considered:
\begin{itemize}
\item[(a)] a coupling hyper-arc, which connects two predecessor trips to one successor trip: $(h_{(v_1, v_2), (v', v'), r})$;
\item[(b)] a coupled-transfer hyper-arc, which connects a coupled predecessor trip to a coupled successor trip: $(h_{(v, v), (v', v'), r})$;
\item (c) a decoupling hyper-arc, which connects a coupled predecessor trip to two successor trips: $(h_{(v, v), (v'_1, v'_2), r})$.
\end{itemize}

\begin{figure}[t!]
\centering
\begin{tikzpicture}[
    trip/.style={circle, draw, minimum size=7mm, inner sep=0pt},
    harc/.style={->, thick},
    label/.style={font=\small}
]

\node[trip] (v1) at (0,1) {$v_1$};
\node[trip] (v2) at (0,-1) {$v_2$};
\node[trip] (vp) at (2.4,0) {$v'$};

\draw[harc] (v1) -- (vp);
\draw[harc] (v2) -- (vp);

\node[label] at (1.2,1.5) {(a) Coupling};

\node[trip] (v) at (4.5,0) {$v$};
\node[trip] (vp2) at (6.9,0) {$v'$};

\draw[harc, bend left=15] (v) to (vp2);
\draw[harc, bend right=15] (v) to (vp2);

\node[label] at (5.7,1.5) {(b) Coupled transfer};

\node[trip] (vd) at (9.0,0) {$v$};
\node[trip] (v1p) at (11.4,1) {$v'_1$};
\node[trip] (v2p) at (11.4,-1) {$v'_2$};

\draw[harc] (vd) -- (v1p);
\draw[harc] (vd) -- (v2p);

\node[label] at (10.2,1.5) {(c) Decoupling};

\node (dotsL) at (0,-3.0) {$\cdots$};
\node[trip] (v1) at (2.3,-3.0) {$v_1$};

\node[trip] (v) at (4.6,-4.3) {$v$};
\node[trip] (vp) at (6.8,-4.3) {$v'$};

\node[trip] (v2) at (9.1,-3) {$v_2$};
\node (dotsR) at (11.4,-3) {$\cdots$};

\node[trip] (vd) at (2.3,-5.6) {$v_d$};
\node[trip] (vdp) at (9.1,-5.6) {$v'_d$};

\node[draw, rectangle, minimum width=1cm, minimum height=0.5cm] (depotL) at (1.0,-5.6) {DEPOT};
\node[draw, rectangle, minimum width=1cm, minimum height=0.5cm] (depotR) at (10.4,-5.6) {DEPOT};

\draw[harc] (dotsL) -- (v1);
\draw[harc] (v2) -- (dotsR);

\draw[harc] (v1) -- (v);
\draw[harc] (vd) -- (v);

\draw[harc, bend left=15] (v) to (vp);
\draw[harc, bend right=15] (v) to (vp);

\draw[harc] (vp) -- (v2);
\draw[harc] (vp) -- (vdp);

\node[label] at (5.7,-2.3) {(d) example EMU circulation starting at depot as $v_d$ and ending at depot as $v_{d'}$};

\end{tikzpicture}
\caption{Illustration of the three hyper-arc types: (a) coupling of two EMUs into a multiple-unit service, (b) transfer of a coupled composition represented by a hyper-arc composed of two parallel arcs, and (c) decoupling into two single-EMU services. In (d), an example of a complete EMU circulation is presented, it starts as the train leaving the depot $v_d$, couples with another EMU to serve trains $v$ and $v'$, and finally decouples to enter the depot as train $v_{d}'$ }
\label{fig:hyperarcs}
\end{figure}

Figure~\ref{fig:hyperarcs}(d) shows a simple example of a complete EMU circulation that starts and ends at a depot. The circulation begins with an EMU departing from depot node $v_d$, then coupling with another EMU to serve trips $v$ and $v'$, and decoupling before one EMU enters the depot through node $v'_d$.
 
Arcs and hyper-arcs have temporal feasibility constraints. The transfer time between trains is bounded from below by $\delta(h_{v,v',r})$ and from above by $\Delta(h_{v,v',r});$ the latter bound limits model complexity and, to some extent, accounts for station capacity. When these bounds are constant across the instance, we simply use $\delta$ and $\Delta$.  Moreover, only a subset of EMU types is allowed to operate in coupled pairs.
Similarly, each railway line can only be served by a predefined subset of EMU types.

 These assumptions simplify the model while retaining practical relevance. 
 We denote by $H$ the set of all feasible arcs and hyper-arcs satisfying temporal, rolling-stock, depot, and coupling requirements. 
 The set includes both trip-to-trip transitions and optional depot service connections.
 
\subsection{ILP encoding}\label{sec::ILP}

\begin{table}[t]
\centering
\caption{Sets used in the formulation.}
\label{tab:sets}
\begin{tabular}{ll}
\toprule
Symbol & Description \\
\midrule
$T$ & Timetable of all trips \\
$V$ & Set of all nodes, these derived from timetable $T$ and service trains \\
$T'$ & Trips that can only be operated by a single EMU \\
$T''$ & Trips allowing single or coupled EMU operation \\
$H$ & Set of feasible arcs and hyper-arcs \\
$H(v)$ & Set of arcs/hyper-arcs covering trip $v$ \\
$R$ & Set of EMU types \\
$D$ & Set of depots \\
$H(v)^{\text{in}}_r$ & Incoming arcs/hyper-arcs for node $v$ and EMU type $r$ \\
$H(v)^{\text{out}}_r$ & Outgoing arcs/hyper-arcs for node $v$ and EMU type $r$ \\
$V_d$ & Set of depot-departure nodes for depot $d$ \\
$V'_d$ & Set of depot-arrival nodes for depot $d$ \\
$H(v_d)_r$ & Feasible arcs originating from depot node $v_d$ for EMU type $r$ \\
$H(v'_d)_r$ & Feasible arcs ending at depot node $v'_d$ for EMU type $r$ \\
$H(t,d)$ & Arcs requiring a driver from depot $d$ at time $t$ \\
\bottomrule
\end{tabular}
\end{table}

The mathematical formulation uses the sets, parameters, and decision variables summarized in Tables~\ref{tab:sets}--\ref{tab:variables}.
Our binary decision variables are assigned to the (hyper)arcs: $x_h \in \{0,1\}$ is equal to one if and only if $h \in H$ is part of the solution. 
The binary integer linear program is the following:
{%
\allowdisplaybreaks
\begin{align}
	\min_{\vec{x}} \quad
	&\alpha \sum_{h \in H} c_h x_h
	+ \sum_{d \in D} \sum_{v_d \in V_d} \sum_{r \in R} \sum_{h \in H(v_d)_r} k'(h) x_h
	\label{eq::obj}
	\\[4pt]
	\text{s.t.} \quad
	&\text{Coverage:}\quad \sum_{h \in H(v)} x_h = 1
	\quad \forall v \in T,
	\label{eq::coverage}
	\\
	&\text{Flow balance:}\quad
	\sum_{h \in H(v)^{\text{in}}_r} k'(h) x_h
	= \sum_{h \in H(v)^{\text{out}}_r} k(h) x_h
	\quad \forall v \in V,\; \forall r \in R,
	\label{eq::conta}
	\\
	&\text{Unique exit:}\quad
	\sum_{r \in R} \sum_{h \in H(v)^{\text{out}}_r} x_h \leq 1
	\quad \forall v \in V,
	\label{eq::contb}
	\\
	&\text{Depot dispatch:}\quad
	0 \leq n(v_d)_r \leq \sum_{h \in H(v_d)_r} k'(h) x_h \leq N(v_d)_r
	\label{eq::depot}
	\\
	&\hspace{9em} \forall d \in D,\; \forall v_d \in V_d,\; \forall r, \notag
	\\
	&\text{Depot return:}\quad
	0 \leq n'(v'_d)_r \leq \sum_{h \in H(v'_d)_r} k(h) x_h \leq N(v'_d)_r
	\label{eq::depot_in}
	\\
	&\hspace{9em} \forall d \in D,\; \forall v'_d \in V'_d,\; \forall r, \notag
	\\
	&\text{Seat limit:}\quad
	\sum_{h \in H:\, p_h > \delta p_h} x_h = 0,
	\label{eq::seats}
	\\
	&\text{Bike limit:}\quad
	\sum_{h \in H:\, b_h > \delta b_h} x_h = 0,
	\label{eq::bikes}
	\\
	&\text{Crew limit:}\quad
	0 \leq a(t,d) \leq \sum_{h \in H(t,d)} k(h) x_h \leq A(t,d)
	\label{eq::drivers_general}
	\\
	&\hspace{9em} \forall d \in D,\; \forall t \in \{t_1,t_2,t_3,t_4,\ldots\}, \notag
	\\
	&\text{Integrality:}\quad
	x_h \in \{0,1\}
	\quad \forall h \in H.
	\label{eq::binary_req}
\end{align}
}%

\begin{table}[t]
\centering
\caption{Parameters used in the formulation.}
\label{tab:parameters}
\begin{tabular}{ll}
\toprule
Symbol & Description \\
\midrule
$\alpha$ & Weighting coefficient in the objective function \\
$k(h)$ & Number of EMUs assigned to target trip(s) of arc/hyper-arc $h$ \\
$k'(h)$ & Number of EMUs assigned to source trip(s) of arc/hyper-arc $h$ \\
$N(v_d)_r$ & Maximum number of EMUs of type $r$ dispatched from depot $d$ \\
$n(v_d)_r$ & Minimum number of EMUs of type $r$ dispatched from depot $d$ \\
$N'(v'_d)_r$ & Maximum number of EMUs of type $r$ returning to depot $d$ \\
$n'(v'_d)_r$ & Minimum number of EMUs of type $r$ returning to depot $d$ \\
$p_h$ & Passenger-seat shortage associated with arc/hyper-arc $h$ \\
$b_h$ & Bicycle-capacity shortage associated with arc/hyper-arc $h$ \\
$\delta p_h$ & Maximum admissible passenger-seat shortage \\
$\delta b_h$ & Maximum admissible bicycle-capacity shortage \\
$c_h$ & Operational cost associated with arc/hyper-arc $h$ \\
$A(t,d)$ & Maximum number of available drivers at depot $d$ and time $t$ \\
$a(t,d)$ & Minimum number of drivers required at depot $d$ and time $t$ \\
\bottomrule
\end{tabular}
\end{table}

\begin{table}[t]
\centering
\caption{Decision variables.}
\label{tab:variables}
\begin{tabular}{ll}
\toprule
Variable & Description \\
\midrule
$x_h \in \{0,1\}$ &
Binary variable equal to 1 if arc/hyper-arc $h$ is selected \\
\bottomrule
\end{tabular}
\end{table}

\paragraph{Objective function}

Eq.~\eqref{eq::obj} combines two operational goals:
\begin{enumerate}
    \item minimizing rolling-stock operating costs;
    \item minimizing the number of deployed EMUs in order to maximize the reserve fleet available for disruption management.
\end{enumerate}

The first term represents operational costs associated with selected arcs and hyper-arcs.
For each arc \(h\), the parameter \(c_h\) includes the fixed daily rolling-stock cost and the distance-dependent operating cost of the corresponding EMU type.

The second term penalizes the number of EMUs dispatched from depots.
Although depot capacities are already enforced through the hard constraints in Eqs.~\eqref{eq::depot} and~\eqref{eq::depot_in}, this additional penalty promotes the preservation of reserve rolling stock. Maintaining a reserve fleet improves operational robustness and increases resilience to future disruptions, such as rolling-stock failures or timetable perturbations.

The weighting coefficient \(\alpha\) controls the trade-off between operational costs and fleet reserve utilization.

\paragraph{Coverage constraint}

Eq.~\eqref{eq::coverage} ensures that each trip is covered once and only once: each trip must be assigned exactly one feasible incoming arc or hyper-arc.

\paragraph{Rolling-stock continuity constraints}

The rolling-stock continuity constraints ensure that EMU flows remain balanced across the timetable and that each trip is followed by at most one feasible continuation.

Eq.~\eqref{eq::conta} enforces flow conservation for each trip node and EMU type by balancing incoming and outgoing EMU flows. The coefficients \(k(h)\) and \(k'(h)\) account for the number of EMUs associated with the target and source trips of a given arc or hyper-arc, respectively, thereby enabling the modelling of coupled EMU operations.

Eq.~\eqref{eq::contb} ensures that each trip has at most one outgoing continuation in the circulation plan. This condition is not guaranteed by Eq.~\eqref{eq::coverage}, which only regulates incoming assignments to trips.

\paragraph{Depot constraints}

These constraints regulate the number of EMUs of each type that may leave and return to individual depots during the planning horizon.

Eq.~\eqref{eq::depot} ensures that the number of EMUs of type \(r\) dispatched from depot \(d\) remains between the prescribed lower and upper bounds \(n(v_d)_r\) and \(N(v_d)_r\), respectively.

Analogously, Eq.~\eqref{eq::depot_in} enforces lower and upper bounds on the number of EMUs of type \(r\) returning to depot \(d\) by the end of the scheduling horizon, using the parameters \(n'(v'_d)_r\) and \(N'(v'_d)_r\).

These constraints may additionally enforce depot balance conditions. In particular, fixing
\[
n(v_d)_r = N(v_d)_r
\quad \text{and} \quad
n'(v'_d)_r = N'(v'_d)_r
\]
imposes an exact number of outgoing and incoming EMUs of type \(r\) for depot \(d\).

\paragraph{Capacity constraints}

These capacity constraints ensure that passenger-seat and bicycle-capacity shortages do not exceed admissible limits.

Eq.~\eqref{eq::seats} eliminates arcs and hyper-arcs for which the passenger-seat shortage exceeds the admissible limit, while Eq.~\eqref{eq::bikes} imposes the analogous restriction for bicycle capacity shortages.

For hyper-arcs representing coupled or split EMU operations involving multiple trips, the corresponding passenger and bicycle capacity conditions must be satisfied simultaneously for all trips connected by the hyper-arc.

\paragraph{Crew availability constraints}

Eq.~\eqref{eq::drivers_general} limits the number of simultaneously operated EMUs requiring drivers from depot \(d\) at time \(t\) to the available driver capacity. The driver availability is evaluated at selected time instants \(t_1,t_2,t_3,\ldots\) over the scheduling horizon. Although the formulation focuses on drivers, the same approach can be directly extended to other crew categories.

A natural extension of the model is to include driver qualifications associated with particular rolling-stock types or railway lines. Let \(L\) denote the set of driver license categories, and let \(H(t,d,l)\) be the set of arcs requiring a driver with license \(l \in L\) from depot \(d\) at time \(t\). The corresponding qualification-dependent driver availability constraint is then given by
\begin{equation}
	\sum_{h \in H(t, d, l)} k(h)x_h
	\leq
	A(t, d, l)
	\quad
	\forall d \in D,\;
	\forall t \in \{t_1,t_2,t_3,\ldots\},\;
	\forall l \in L.
	\label{eq:drivers_particular}
\end{equation}
Analogous lower-bound constraints based on \(a(t,d,l)\) may also be introduced if required.

To evaluate the problem's size, we assume the worst-case limit for number of arcs per node:
\begin{equation}
|H(v)| \leq |T|^2 |R|
\label{eq::size_H}
\end{equation}
where $|R|$ is the number of types of EMUs (this assumes for simplicity $|T|$ = $|V|$, but if we want to be precise, it is enough to replace $T$ by $V$).
Here, the arc to the given train can be from all other trains, with all possible EMU types, and the quadratic factor accounts for all possible pairs of source trips in coupled-EMU arcs. 
Regarding scaling, in the worst case, the problem size (i.e., the number of binary variables $|\vec{x}|$, equal to the number of arcs and hyper-arcs) is
\begin{equation}
	|\vec{x}| \leq  |T| \cdot |T|^2 |R|
\end{equation}
However, since the coupling of EMUs is allowed only in a predefined subset of trips $T'' \subset T$, the actual problem is smaller. In the worst case we obtain
\begin{equation}
	|\vec{x}| \leq  |T'|^2 \cdot |R| + |T''|^3 \cdot |R| 
	\label{eq::ilp_scaling}
\end{equation}

\subsection{QUBO encoding}

As discussed before, we have selected the QUBO-based approach as it is native to quantum annealing hardware, which, at the current state of the art, can handle relatively large optimization problems.
Hence, for quantum and quantum-inspired approaches, we encode our problem as Quadratic Unconstrained Binary Optimization (QUBO). In this implementation, we treat all constraints as soft with large penalties and then post‑filter infeasible solutions, rather than trying to guarantee feasibility by construction. We follow a standard ILP--to--QUBO transformation as in~\cite{glover2018tutorial}.

The general QUBO form is as follows
\begin{equation}
\min_{\vec{y}} \vec{y}^{\intercal} Q \vec{y}
\label{eq::qubo}
\end{equation}
where $\vec{y}$ is a vector of binary variables, $\forall_i\,  y_i \in \{0,1\}$.
In our case, $\vec{y}$ collects both the decision variables $x_h$ from Section~\ref{sec::ILP} and additional auxiliary slack variables $\forall_i\,  s_i \in \{0,1\}$. We transform the problem defined by Eqs.~\eqref{eq::obj}–\eqref{eq::binary_req} into the following quadratic form:

\begin{equation}
 \min_{\vec{y}} \left( f(\vec{y}) + P_1(\vec{y}) + P_2(\vec{y}) + P_3(\vec{y}) + P_4(\vec{y}) + P_5(\vec{y}) \right)
\end{equation}
where the objective is:
\begin{equation}
   f(\vec{y}) =
\sum_{h \in H} \alpha c_h x_h
+ \sum_{d \in D} \sum_{v_d \in V_d} \sum_r \sum_{h \in H(v_d)_r} k'(h) x_h.
\label{eq::qubo_objective}
\end{equation}
In constraint terms $P_1, \ldots P_5$ we use penalty coefficients: $\lambda_1, \ldots \lambda_5$ are penalty coefficients used to convert the constrained problem into an unconstrained QUBO. 
The coverage constraint from Eq.~\eqref{eq::coverage} yields following QUBO terms
\begin{equation}
    P_1(\vec{y}) = \lambda_1 \sum_{v \in T}
\left( \sum_{h \in H(v)} x_h - 1 \right)^2.
\label{eq::qubo_coverage}
\end{equation}
 The rolling-stock continuity constraints from Eqs.~\eqref{eq::conta} and~\eqref{eq::contb} yield
\begin{align}
   P_2(\vec{y}) &= \lambda_2 \Bigg[
\sum_{v \in V} \sum_{r \in R}
\left(
\sum_{h \in H(v)^{\text{in}}_r} k'(h) x_h
-
\sum_{h \in H(v)^{\text{out}}_r} k(h) x_h
\right)^2
\notag\\
&\qquad\qquad
+ \sum_{v \in V}
\left(
\sum_{r \in R} \sum_{h \in H(v)^{\text{out}}_r} x_h - s
\right)^2
\Bigg]
\label{eq::continuity}
\end{align}
The rolling stock limits in depots from Eq.~\eqref{eq::depot} and~\eqref{eq::depot_in}, yield:
\begin{align}
    P_3(\vec{y}) &= \lambda_3 \Bigg[
\sum_{d \in D} \sum_{v_d \in V_d} \sum_r
\left(
\sum_{h \in H(v_d)_r} k'(h) x_h
- n(v_d)_r
- \sum_{k=1}^{N(v,d)-n(v,d)} s'_k
\right)^2
\notag\\
&\qquad\qquad
+ \sum_{d \in D} \sum_{v'_d \in V'_d} \sum_r
\left(
\sum_{h \in H(v'_d)_r} k(h) x_h
- n'(v'_d)_r
- \sum_{k=1}^{N'(v',d)-n'(v',d)} s''_k
\right)^2
\Bigg]
\label{eq::qubo_depot}
\end{align}
Capacity constraints for passengers and bicycles imposed in Eqs.~\eqref{eq::seats}–\eqref{eq::bikes}, yield:
\begin{align}
   P_4(\vec{y}) = \lambda_4  \left( \sum_{h \in H:\, p_h > \delta p_h} x_h + \sum_{h \in H:\, b_h > \delta b_h} x_h \right)
\label{eq::qubo_seats}
\end{align}
Finally the crew constraint in Eq.~\eqref{eq::drivers_general} yields:
\begin{equation}
   P_5(\vec{y}) =  \lambda_5 \sum_{d \in D} \sum_{t \in \{t_1,t_2,t_3,t_4,\ldots\}}
\left(
\sum_{h \in H(t,d)} k(h) x_h
- a(t,d)
- \sum_{k=1}^{A(t,d)-a(t,d)} s'''_k
\right)^2.
\label{eq:drivers_qubo}
\end{equation}
In Eqs.~\eqref{eq::continuity}, \eqref{eq::qubo_depot} and \eqref{eq:drivers_qubo}  we use slack variables $s, s', s'', s''' \in \{0,1\}$ to reflect inequalities in Eqs.~\eqref{eq::conta}, \eqref{eq::contb}, \eqref{eq::depot}, \eqref{eq::depot_in} and~\eqref{eq::drivers_general}. 



Optionally, the constraint in Eq.~\eqref{eq:drivers_particular} can be encoded in a  similar way using additional slack variables.

Using the QUBO encoding, the overhead in the number of terms (compared with the ILP scaling in Eq.~\eqref{eq::ilp_scaling}) is entirely due to the slack variables, since all ILP decision variables are already binary:
\begin{equation}
	|\vec{y}| = |\vec{x}| + \# \text{slack vars} 
\end{equation}

Denoting total number of EMUs as $\# EMUs$, $|T|$ as number of trains, and $|D|$ as number of stations - the number of slack variables can be bounded from above by $2 N$ (where $N = \max_{v,d} N(v,d) \leq \# EMUs$ is the number of rolling stock in depots) from Eq.~\eqref{eq::qubo_depot} and $\# crew \cdot \{t_1, t_2, t_3, t_4, \ldots\}$ from Eq.~\eqref{eq:drivers_qubo}, and $ |T| \cdot |D|$ from  Eq.~\eqref{eq::continuity} namely:
\begin{equation}
	\# \text{slack vars} \leq 2 \# EMUs + \# crew \cdot \{t_1, t_2, t_3, t_4, \ldots\} +  |T| \cdot |D|
\end{equation}
In practice, in our instances the number of slack variables is small compared with the number of decision variables.

Regarding the number of QUBO terms, we obtain the following bounds:
\begin{itemize}
    \item from Eq.~\eqref{eq::qubo_coverage} if we bound $|H(v)|$ via Eq.~\eqref{eq::size_H}, we have:
    \begin{equation}
    \# terms \leq |T| |H(v)|^2 \leq |T|^5 |R|^2 
    \end{equation}
    \item from Eq.~\eqref{eq::continuity} if we bound $\sum_{r \in R} |H(v)_r^{in}|, \sum_{r \in R}  |H(v)_r^{out}|  \leq  2  |R| |T|^2$ (the square is from possible joining) and $|V| \leq 2 \#( EMU types) |T| $ then:
    \begin{equation}
    \# terms \leq 2 |R| |T| \left(|R| |T|^2\right)^2
    \end{equation}
    In practice, for each node $v$ and EMU type $r$ we limit the number of trains that arcs can start from or lead to, so that $|H(v)|$, $|H(v)^{\text{in}}_r|$ and $|H(v)^{\text{out}}_r|$ can be reduced by the parameters $\delta$ and $\Delta$. 
    \item From Eq~\eqref{eq::qubo_depot}, if we assume that the number of trips leaving depots is at most $|T|$, we obtain
    \begin{equation}
    \# terms \leq 2 |R| |T| \left(2 |R| |T| + N\right)^2
    \end{equation}
    where $N = \max_{v,d} N(v,d) \leq \# EMUs$ is the number of rolling stock in depots.
    \item Eq.~\eqref{eq::qubo_seats} yields a bound of the same order as
    Eq.~\eqref{eq::seats}~\eqref{eq::bikes}.
    \item From Eq.~\eqref{eq:drivers_qubo}
    \begin{equation}
    \# terms \leq |D| |\{t_1, t_2, t_3, t_4, \ldots \}| \left(2 |R| |T| + A\right)^2
    \end{equation}
    where $A = \max_{t,d} A(t,d)$ is the number of drivers.
\end{itemize}

Concluding, the number of QUBO terms scales in the worst case as $|R|^3$ and $|T|^5$. This is to be compared with the scaling of the linear formulation in Eq.~\eqref{eq::size_H}: there the scaling in rolling stock types is linear. As in practical instances $|T|$ is order of $100$, this suggests that the QUBO approach will be less practical in case of large instances. Yet it is conceptually interesting as it is one of the native input formats of quantum devices.

\subsection{Toy example}\label{sec::toy}

Let us now demonstrate in detail the actual behavior of our model on a toy example that can be followed in detail.
The toy model is defined by the timetable $T = \{ v_1, v_2, v_3\}$ where $v_1$ and $v_2$ are trains that go $A \rightarrow B$, and $v_3$ goes $B \rightarrow A$. The timetable is arranged so that any rolling stock that arrives at $B$ can either remain there or return to $A$. Two EMU types $R = \{r_1, r_2\}$ are available at the depot at station $A$ throughout the time horizon, with fleet sizes $N(v_d)_{r_1} = 2, N(v_d)_{r_2} = 1$ with the seats capacity $p_1 = 70$, $p_2 = 110$. There is also an optional service train $v_4$ going from $B$ to $A$ (which can be served only by type $r_1$). We assume that $A = 2$ drivers are available throughout the horizon; they can drive any of the two EMUs and take any of the trips. Finally, we do not consider bicycles in this example.

We assume the following costs:
\begin{itemize}
 \item operating costs of EMUs (the same on both trips): $r_1 - 70$, $r_2 - 110$ (we assume these costs are additive);
 \item expected numbers of passengers on the trips: $v_1 - 70$, $v_2 - 60$, $v_3 - 100$,
 \end{itemize}
For this example we assume that the number of passengers is constant over each trip and that the seat shortage cannot exceed $\delta p_{h} = 10$ for a single EMU and $\delta p_{h} = 20$ for coupled EMUs.
Let us also assume that EMUs can not be coupled at station $A$, but they can be coupled at station $B$ and only type $r_1$ is allowed to form coupled compositions.

The hyper-graph of the problem is composed of the following nodes: $V = \{v_{d}, v_{1}, v_{2}, v_{3}, v_4\}$ where $v_{d}$ represents service trains from the depot that are not included in timetable and $v_4$ represents the optional service train. Let 
\begin{align}
&H = \{h_{v_{d}, v_{1}, r_1}, h_{v_{d}, v_{1}, r_2}, h_{v_{d}, v_{2}, r_1}, h_{v_{d}, v_{2}, r_2}, h_{v_1, v_{3}, r_1}, h_{v_1, v_{3}, r_2}, h_{v_1, v_4, r_1}, \nonumber \\ &h_{v_2, v_{3}, r_1}, h_{v_2, v_{3}, r_2}, h_{v_2, v_4, r_1} , h_{(v_1, v_2), (v_{3}, v_{3}), r_1}\} \nonumber \\
 &=> \{ x_0, x_1, \ldots, x_8, x_9, x_{10} \}
	 \label{eq:set_H}
\end{align}
where the variable $x_{10}$ corresponds to the hyper-arc $h_{(v_1, v_2), (v_{3}, v_{3}), r_1}$ that represents coupling two EMUs at station $B$. 

We then have
\begin{align}	H(v_1)^{\text{in}} = \big\{ 
	h_{v_{d}, v_{1}, r_1},  h_{v_{d}, v_{1}, r_2} \} , \  H(v_2)^{\text{in}} = \big\{
	h_{v_{d}, v_{2}, r_1},  h_{v_{d}, v_{2}, r_2} \}  => 	\{0,1\} \ \{2,3\}
	\label{eq:set_Hin}
\end{align}
and
\begin{align}
	H(v_1)^{\text{out}} = \big\{ &
	h_{v_1, v_{3}, r_1},  h_{v_1, v_{3}, r_2}, h_{v_1, v_{4}, r_1},  h_{(v_1, v_2), (v_{3}, v_{3}), r_1} \} , \nonumber \\  H(v_2)^{\text{out}} = \big\{&
	h_{v_2, v_{3}, r_1},  h_{v_2, v_{3}, r_2}, h_{v_2, v_{4}, r_1}, h_{(v_1, v_2), (v_{3}, v_{3}), r_1} \} \nonumber \\
	 &=> 	\{4,5,6,10\} \ \{7,8,9,10\}
	\label{eq:set_Hout}
\end{align}
while
\begin{equation}
\begin{split}
H(v_{d})_{r=1} = \{h_{v_{d}, v_{1}, r_1}, h_{v_{d}, v_{2}, r_1} \} \text{ and } H(v_{d})_{r=2} =  \\ \{h_{v_{d}, v_{1}, r_2}, h_{v_{d}, v_{2}, r_2} \} => 	\{0,2\} \ \{1,3\}
\end{split}
\end{equation}
and
\begin{align}	H(v_1) = \big\{ &
	h_{v_{d}, v_{1}, r_1},  h_{v_{d}, v_{1}, r_2} \} , \  H(v_2) = \big\{
	h_{v_{d}, v_{2}, r_1},  h_{v_{d}, v_{2}, r_2} \}  => 	\{0,1\} \ \{2,3\} \nonumber \\
	H(v_3) = \big\{ &
	h_{v_1, v_{3}, r_1},  h_{v_1, v_{3}, r_2}, h_{v_2, v_{3}, r_1},  h_{v_2, v_{3}, r_2}, h_{(v_1, v_2), (v_{3}, v_{3}), r_1} \} => 	\{4,5,7,8,10\}.
	\label{eq:set_Htau}
\end{align}

\paragraph{ILP formulation} 

Assume $\alpha = 0.01$. The objective becomes:
\begin{align}
	& \min_{\vec{x}} \left( \alpha \sum_{h \in H}  c_h x_h  +  \sum_{r \in \{r_1, r_2\}} \sum_{h \in H(v_{d})_r} k(h) x_h \right) \nonumber \\
	& = \bigg( 0.7 x_0 + 1.1 x_1 + 0.7 x_2 + 1.1 x_3 + 0.7 x_4 + \nonumber \\
&\quad 1.1 x_5 + 0.7 x_6 + 0.7 x_7 + 1.1 x_8 + 0.7 x_9 + 1.4 x_{10}   \bigg) + \nonumber \\ & \left(
	1 x_0 + 1 x_1 + 1 x_2 + 1 x_3
	 \right)
\end{align}
The coverage constraint in Eq.~\eqref{eq::coverage} would be:
\begin{align}
\sum_{h \in H(v)} x_h = 1 &\quad \forall v \in T \nonumber\\
x_0 + x_1 = 1  \ \ x_2 + x_3 = 1 \ \ &x_4 + x_5 + x_7 + x_8 + x_{10} = 1
\end{align}
The rolling-stock continuity constraints in Eqs.~\eqref{eq::conta} and~\eqref{eq::contb} are:
\begin{align}
\sum_{h \in H(v)^{\text{in}}_r} &k'(h) x_h = \sum_{h \in H(v)^{\text{out}}_r} k(h) x_h \quad \forall_{v \in \{v_1, v_2\} r \in \{r_1, r_2\}} \nonumber \\
&x_0 = x_4 + x_6 + x_{10} \text{ and }  x_1 = x_5 \ (\text{for } v_1) \nonumber \\
&x_2 = x_7 + x_{10} + x_9 \text{ and }  x_3 = x_8 \ (\text{for } v_2)
\end{align}
and
\begin{align}
 \sum_{r \in \{r_1, r_2\}} \sum_{h \in H(v)^{\text{out}}_r}  x_h &\leq 1 \quad \forall_{v \in \{v_1, v_2\}} \nonumber \\
 x_{10} + x_4 + x_5 + x_6 \leq 1 \quad &x_{10} + x_7 + x_8 + x_9 \leq 1
 \end{align}
Depot constraint Eq.~\eqref{eq::depot} (we drop the return constraints in Eq.~\eqref{eq::depot_in}) reduce to
\begin{align}
	\sum_{h \in H(v_d)_r}  k(h) x_h & \leq N(v_d)_r \quad \forall_{v_d \in \{v_{d}\}}  \forall_{r \in \{r_1, r_2\}} 
	\nonumber \\
	x_0 + x_2 & \leq 2 \quad x_1 + x_3 \leq 1
\end{align}
Concerning capacity (seats) constraint
\begin{align}
&\sum_{h \in H: p_{h}  > \delta p_{h}} x_h = 0 \nonumber \\  
& \quad x_4 + x_7 = 0
\end{align}
Other capacity constraints for this model are trivial, hence not included.
For the crew availability (driver's) constraint, let us assume that
\begin{align}
 H(t_1, d) &= \{h_{v_{d}, v_{1}, r_1}, h_{v_{d}, v_{1}, r_2} \}  \ \ A(t_1,d)  =  2 \\
 H(t_2, d) &= \{ h_{v_1, v_{3}, r_1}, h_{v_1, v_{3}, r_2}, h_{v_1, v_{4}, r_1}, h_{v_2, v_{3}, r_1} h_{v_2, v_{3}, r_2}, h_{v_2, v_{4}, r_1}, \nonumber \\
&\quad h_{(v_1, v_2), (v_{3}, v_{3}), r_1} \}  \ \ A(t_2,d)  =  2 
\end{align}
yielding
\begin{align}
	\sum_{h \in H(t, d)} x_h  \leq A(t,d) &\quad \forall_{d \in D} \forall_{(t) \in \{t_1, t_2\}} \nonumber \\ 
	x_0 + &x_1 \leq 2 \nonumber \\
	x_4 + x_5 + x_6 + x_7 &+ x_8 + x_9 + x_{10}  \leq 2
\end{align}

\begin{figure}[t!]
	\centering
	\begin{tikzpicture}[>=Stealth, node distance=2cm]
		
		\node[circle, draw, minimum size=1cm] (Vd) at (-1,1) {$V_d$};
		\node[circle, draw, minimum size=1cm] (V3) at (6,1) {$V_3$};
		\node[circle, draw, minimum size=1cm] (V2) at (2,0) {$V_2$};
		\node[circle, draw, minimum size=1cm] (V1) at (2,2) {$V_1$};
		\node[circle, draw, minimum size=1cm] (V4) at (6,2.5) {$V_4$};
		
		\draw[->, black, thick] (Vd) -- (V2) node[midway, below, black] {$h_{v_d, v_2, r_1}$}; 
		\draw[->, darkgreen, thick] (V1) -- (V3);
		\draw[->, black, thick] (Vd) -- (V1)node[midway, above, yshift=2pt, black] {$h_{v_d, v_1, r_1}$}; 
		\draw[->, darkgreen, thick] (V2) -- (V3) node[midway, above, yshift=5pt, darkgreen] {$h_{(v_1, v_2), (v_3, v_3), r_1}$}; 
		
	\end{tikzpicture}
	\caption{Illustration of the network structure of the optimal solution of the toy problem. Black arrows represent arcs, while green arrows represent the single hyper-arc.}
	\label{fig:network_ground_solution}
\end{figure}
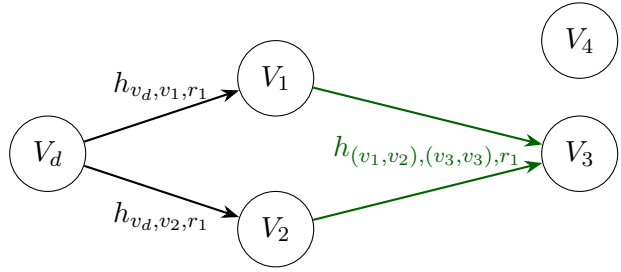

We obtain the optimal solution $x_0 = 1, x_2 = 1, x_{10} = 1$ with $f(\vec{x}) = 4.80$. The corresponding network structure is shown in
Fig.~\ref{fig:network_ground_solution}. We also identify two feasible excited solutions:
\begin{itemize}
	\item $x_0 = 1, x_3 = 1, x_6 = 1, x_8 = 1$ with $f(\vec{x}) = 5.6$
	\item $x_1 = 1, x_2 = 1, x_5 = 1, x_9 = 1$ with $f(\vec{x}) = 5.6$
\end{itemize}

\paragraph{QUBO formulation}

We convert the ILP problem into the QUBO using the penalty method, i.e., by adding quadratic penalty terms corresponding to Eqs.~\eqref{eq::qubo_objective}-~\eqref{eq::qubo_seats} and Eq.~\eqref{eq:drivers_qubo}. Then the resulting QUBO objective is:
\begin{align}
	& \min_{\vec{x}}:\nonumber \\
    & 1.7 x_0 + 2.1 x_1 + 1.7 x_2 + 2.1 x_3 + 0.7 x_4 + 1.1 x_5 + 0.7 x_6 + 0.7 x_7 + 1.1 x_8  + \nonumber \\
    &\quad 0.7 x_9  +1.4 x_{10} +\\  
	& \lambda_1  \left( \left(x_0 + x_1  - 1 \right)^2 + \left(x_2 + x_3  - 1 \right)^2 + \left(x_4 + x_5 + x_7 + x_8 + x_{10}  - 1 \right)^2\right)+  \\
	&\lambda_2 \bigg(  \left( x_4 + x_6 + x_{10} - x_0 \right)^2 + \left( x_5 - x_1 \right)^2 + \left( x_7 + x_9 + x_{10} - x_2 \right)^2  + \left( x_8 - x_3 \right)^2  + \nonumber \\ &( x_4 + x_5 + x_6 + x_{10} - s_{11})^2 + ( x_7 + x_8 + x_9 + x_{10} - s_{12})^2 \bigg) + 
	\\ 
	&\lambda_3 \left( \left( x_0 + x_2 -  s_{13} - s_{14} \right)^2  +\left( x_1 + x_3 -  s_{15} \right)^2 \right) +
	\\
	&\lambda_4 \left( x_4 + x_7  \right) +
	\\
	&\lambda_5 \left(  \left( x_0 + x_1 -  s_{16} - s_{17} \right)^2 + \left( x_4 + x_5 + x_6 + x_7 + x_8 + x_9 + x_{10} - s_{18} - s_{19} \right)^2 \right)
\end{align}
where $s_i \in \{0,1\}$ are slack variables used to encode the inequality
constraints. 

An interesting feature of this formulation is that, if we set $\lambda_4$ small relative to the other penalty coefficients, solutions that violate only the seat capacity constraint (i.e., that corresponds to overcrowded trains) may still appear among low-energy states, while all other constraints remain satisfied.

In the toy example, we used the D-Wave simulator with $\lambda_1 = \lambda_2 = \lambda_3 = \lambda_4 = \lambda_5 = 100$  for the reason mentioned above. With this choice we obtain the following set of low-energy solutions, both feasible and non-feasible:
 \begin{itemize}
 	\item Optimum: $x_0 = 1, x_2 = 1, x_{10} = 1$ and $f(\vec{x}) = 4.8$
 	\item $x_0 = 1, x_3 = 1, x_6 = 1, x_8 = 1$ and $f(\vec{x}) = 5.6$
 	\item $x_1 = 1, x_2 = 1, x_5 = 1, x_9 = 1$ and $f(\vec{x}) = 5.6$
 	 \item Not feasible (capacity) $x_0 = 1, x_2 = 1, x_6 = 1, x_7 = 1$
 	\item Not feasible (capacity) $x_1 = 1, x_2 = 1, x_4 = 1, x_9 = 1$ 
 \end{itemize}

\section{Computational Results}\label{sec::solution}

The goal is to obtain feasible solutions with practical significance, such that the dispatcher will be able to use them in practice. Additionally, having access to a spectrum of feasible solutions allows the dispatcher to choose one, especially taking into account non-encoded conditions. 

\subsection{Solution approaches}

We consider two encodings of the rolling-stock circulation problem:
\begin{itemize}
\item an ILP formulation solved with the SCIP solver~\cite{BolusaniEtal2024ZR}, and
\item a QUBO formulation suitable for quantum annealers and physics-inspired heuristics.
\end{itemize}

In order to solve ILPs, we opt for the SCIP~\cite{BolusaniEtal2024ZR} (version 9.2.1.) as it is probably the most efficient non-commercial linear integer programming solver at the moment. According to our experience, this solver could be used even in production in the studied railway environment.

As for QUBOs, they have a relevant literature also in classical optimization (see e.g. the monograph edited by Punnen~\cite{PunnenQUBO}), and they can also be solved with classical solvers. The aforementioned overhead in the number of variables of the QUBO model suggests that our direct reformulation of the ILP to QUBO will not be practical for bigger instances. Indeed, our aim is to explore the behavior of quantum hardware and physics-based QUBO heuristics rather than to consider this as a practical solution for big instances.

Quantum annealers, such as DWave which is used in the present contribution exploit the equivalence of QUBOs with Ising models prevalently known in physics. A quantum annealer actually realizes a configurable physical setup whose lowest-energy state gives an optimal configuration for QUBOs\cite{albash2018adiabatic}. A quantum annealer is not a universal solver of QUBOs, however, in the case of certain instances, it can potentially be more efficient than classical algorithms. It is heuristic by nature: it provides a set of results which are likely be close to optimal and potentially include the optimal solution. It should be also stressed that quantum annealers have a particular topology, meaning that a general QUBO can only be solved on them after having the respective Ising model's adjacency graph \emph{embedded} to the graph implemented by the hardware as an induced subgraph. This eventually requires multiple spins to be coupled to represent a single spin (i.e. variable) of the problem.

A physics-based alternative direction in solving QUBOs is the application of quantum-inspired algorithms~\cite{oshiyama2022benchmark}.
Following this path, we apply the new VeloxQ solver~\cite{tuziemski2025veloxq}, which appears to be effective for large, sparse QUBOs (with a limited number of connections)~\cite{Robertson_2025, hanussek2025solvingquantuminspireddynamicsquantum, pawlowski2025closingquantumclassicalscalinggap, tuziemski2025recentquantumruntimedisadvantages}. This heuristic solver takes a QUBO or Ising instance and returns a series of solutions intended to originate from the problem's low-energy spectrum (hopefully including at least one feasible solution).
VeloxQ is a physics-inspired classical solver for QUBO/Ising problems that mimics energy-minimization dynamics without requiring quantum hardware and achieves high scalability on standard CPUs/GPUs.
Compared to D-Wave quantum annealing and simulated bifurcation, it delivers comparable or superior solution quality and runtime, especially for large problem instances. As the solver does not need embedding, it is efficient for problems with arbitrary graph structure.

\subsection{Case studies and experimental setup}

The proposed model was evaluated on a collection of railway circulation-planning instances derived from the real timetable of Koleje Śląskie (Silesian Railways, Poland) from December 2024.  
The instances correspond to daily rolling stock schedules on networks of gradually increasing size: from the toy example with $3$ trips to practical regional networks with over $400$ trips and up to $3.8 \cdot 10^5$ ILP variables.
The experiments were designed to evaluate:
\begin{itemize}
    \item scalability of the ILP and QUBO formulations,
    \item the impact of coupled-EMU operations,
    \item sensitivity to increased passenger demand,
    \item the influence of the transfer-time parameter $\Delta$,
    \item practical applicability of quantum and quantum-inspired solvers.
\end{itemize}

\begin{figure}[t!]
    \centering
    \includegraphics[width=0.9\linewidth]{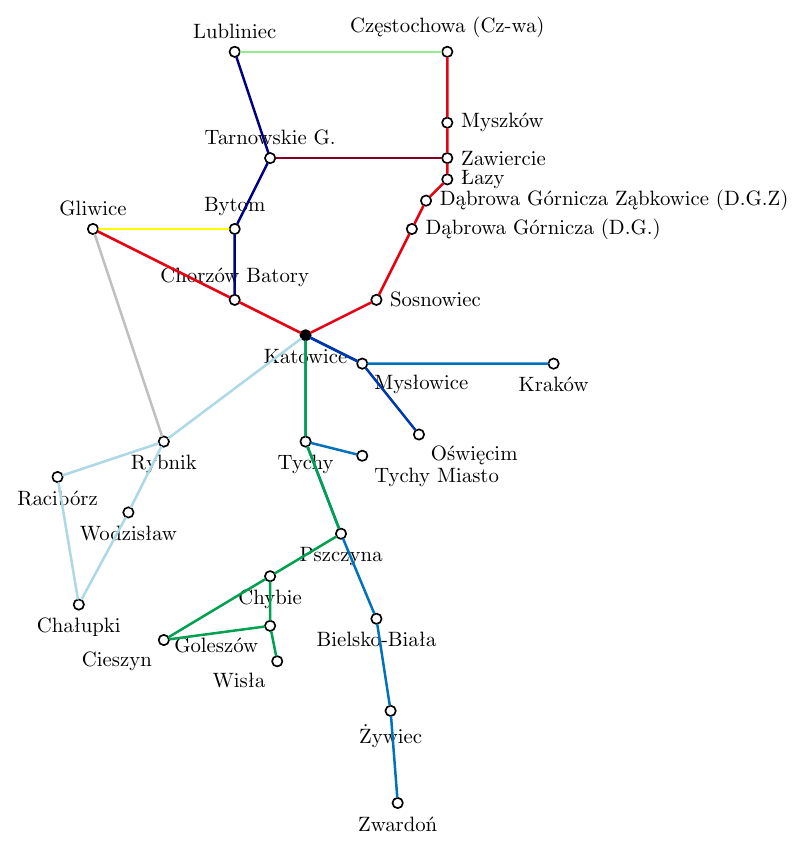}
    \caption{Schematic topology of the network, for the more detailed one see \href{https://www.kolejeslaskie.com/rozklad_jazdy/schemat-linii-komunikacyjnych/}{https://www.kolejeslaskie.com/rozklad\_jazdy/schemat-linii-komunikacyjnych/}}
    \label{fig:enetwork-topology}
\end{figure}
Beyond the toy model from Section~\ref{sec::toy}, the smallest practical instance (instances~$1$ and $1$ a)) concerns the daily traffic between two neighbouring stations, Gliwice and Bytom. Subsequent instances gradually increase the size of the considered network, namely:
\begin{itemize}
    \item $2$ $2$ a) --- Gliwice--Bytom, Gliwice--Rybnik and Rybnik--Racibórz,
    \item $3, 4$, $4$ a) --- Gliwice--Częstochowa and intermediate stations (case 3 has limited $\Delta$ parameter values),
    \item $5, 5$ a) --- as in cases $4$, $4$ a) and Gliwice--Bytom and Gliwice--Rybnik 
    \item $6, 6$ a) --- as in cases $5$, $5$ a) and Częstochowa--Tarnowskie Góry.
\end{itemize}
The largest instances ($7$, $7$ a) and $8$, $8$ a) concern traffic across almost the entire network, with some limitations due to data availability (Czestochowa--Lubliniec) and limited train service caused by engineering works (Żywiec--Zwardoń). Cases $8$ and $8$ a) use enlarged $\Delta$ parameter value. 

For most of the above base instances there exist corresponding variants with increased passenger and bicycle demand, denoted by 'a)'. Finally, since we consider daily rolling stock circulation plans, we consider as depots central stations (such as Gliwice, Katowice, and Częstochowa) as well as terminal (edge) stations, where trains start their routes in the morning and/or end in the evening. This is a new interesting aspect in our model, enabling the use of daily schedules in the practical application of rescheduling or as part of a larger multi-day model.
The specifics and sizes of all instances are presented in Tab.~\ref{tab::instances}.

\begin{table}[ht!]
\centering
  \setlength{\tabcolsep}{2.8pt}
\begin{tabular}{|l|l|l|l|l|l|l|l|l|}
	\hline
	inst. & $|T|$ & $|T'|/ |T''|$& $|D|$ & $|R|$ & $\delta$ & $\Delta$ & \begin{tabular}{c} \#  ILP \\ vars \end{tabular} & \begin{tabular}{c} \# QUBO \\ vars / terms \end{tabular} \\ \hline
	toy  &   3 & 2/1 &   1  & 2 &10 &60 & $11$ & $20$/ $88$ \\ \hline
	1 & 30 & 30/0 & 1&1 & 5 & 60 & $33$ & $42$/ $93$ \\ \hline
	1a & 30 & 0/30 & 1 & 1 & 5 & 60 & $66$ & $103$ / $371$ \\ \hline
	2 & 51 & 51/0 & 3 & 1 & 5 & 60 & $74$ & $103$ / $309$ \\ \hline
	2a & 51 & 21/30 & 3 & 1 & 5 & 60 & $120$ & $172$ / $800$ \\ \hline
        3 & 78 & 30/48 & 6 & 5 & 5 & 60 - 150 & $1599$ & $1756$ / $87121$ \\ \hline
	4 & 78 & 30/48 & 6 & 5 & 5 & 300 & $6238$ & $6295$ / $1.8 \cdot 10^6$ \\ \hline
	4a & 78 & 30/48 & 6 & 5 & 5 & 300 & $6238$ & $6295$ / $1.8 \cdot 10^6$ \\ \hline
	5 & 126 & 78/48 & 7 & 11 & 5 & 300 & $4.8 \cdot 10^4$ & $4.8 \cdot 10^4$ / $8.5 \cdot 10^7$ \\ \hline
	5a & 126 & 78/48 & 7 & 11 & 5 & 300 & $4.8 \cdot 10^4$ & $4.8 \cdot 10^4$ / $8.5 \cdot 10^7$ \\ \hline

 6 & 143 & 95/48 & 8 & 11 & 5 & 300 & $5.2 \cdot 10^4$ & $5.2 \cdot 10^4$ / $9.8 \cdot 10^7$ \\ \hline
	6a & 143 & 95/48 & 8 & 11 & 5 & 300 & $5.2 \cdot 10^4$ & $5.2 \cdot 10^4$ / $9.8 \cdot 10^7$ \\ \hline
	7 & 404 & 356/48 & 19 & 11 & 5 & 300 & $8.2 \cdot 10^4$ & $8.3 \cdot 10^4$ / $1.5 \cdot 10^8$ \\ \hline
	7a & 404 & 356/48 & 19 & 11 & 5 & 300 & $8.2 \cdot 10^4$ & $8.3 \cdot 10^4$ / $1.5 \cdot 10^8$ \\ \hline
	8 & 404 & 356/48 & 19 & 11 & 5 & 960 & $3.8 \cdot 10^5$ & \text{n.a.}
 \\ \hline
	8a & 404 & 356/48 & 19 & 11 & 5 & 960 & $3.8 \cdot 10^5$ & \text{n.a.} \\ \hline
\end{tabular}
	\caption{Instance descriptions. Instances marked by a) have increased passenger and bicycle demand. For large instances number of QUBO variables is similar to the number of ILP variables (the impact of slack variables is limited), but the size of the QUBO in terms of its terms grows rapidly. The largest QUBOs could not be constructed due to the memory limits of the $62$ GB RAM. (Recall that $T'$ are trips that can be served only by the single EMU, while $ T''$ are trips that can be served by either single or multiple EMUs.)}
  	\label{tab::instances}
\end{table}

\subsection{ILP results}
\begin{figure}[t!]
    \subfloat[]{\includegraphics[width=0.5\textwidth]{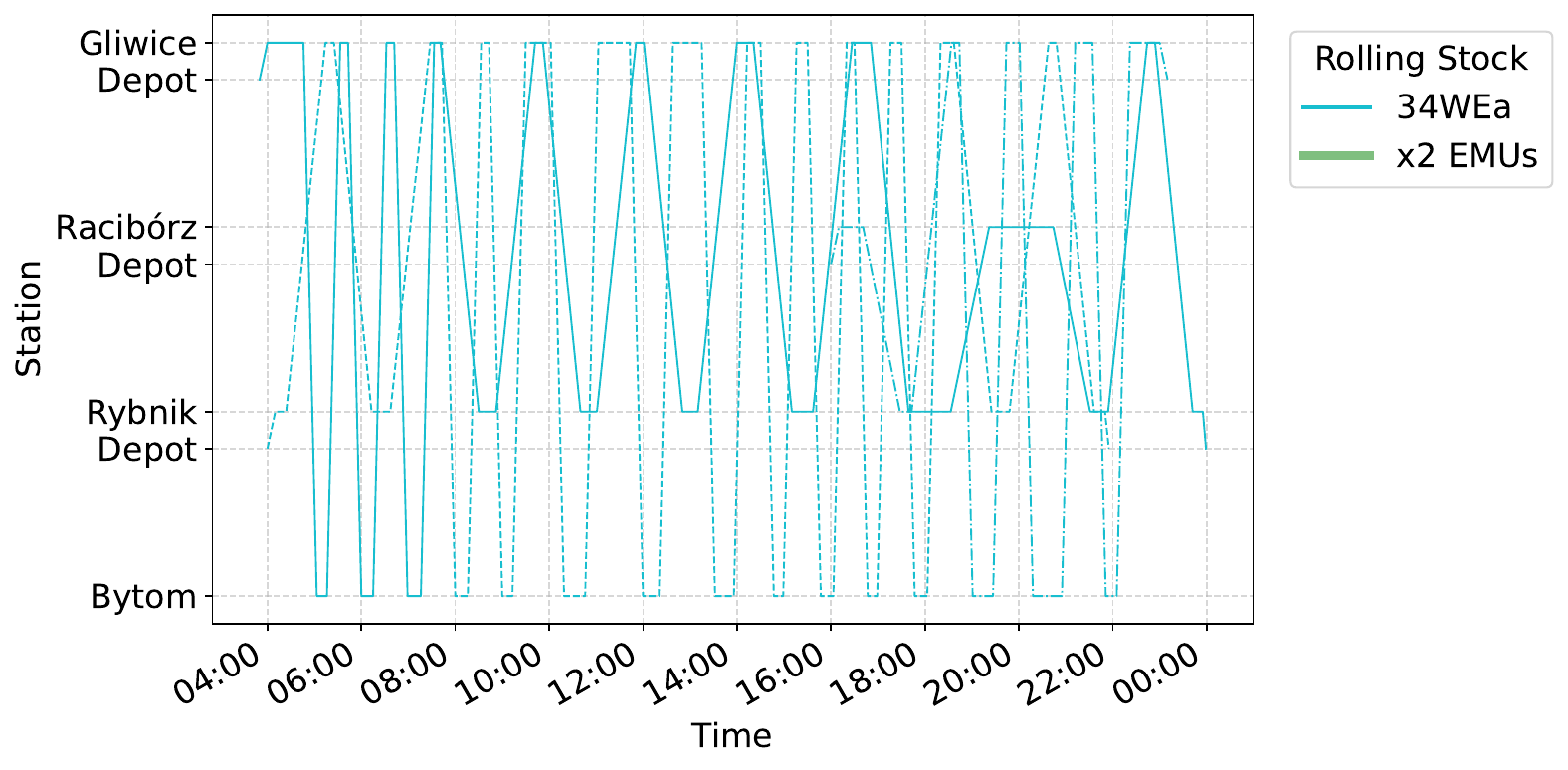}}%
    \subfloat[]{\includegraphics[width=0.5\textwidth]{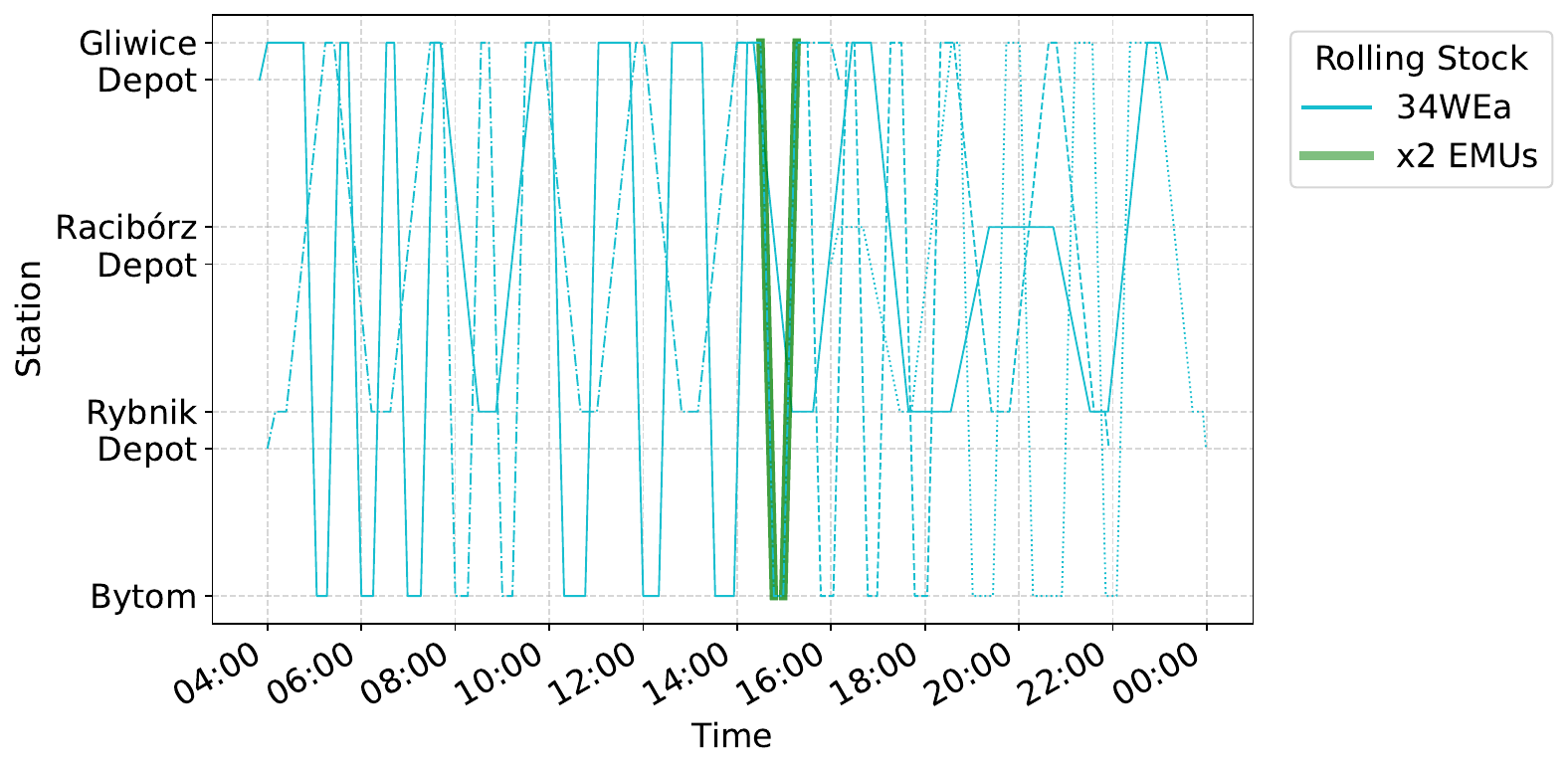}}
    \caption{Train diagrams of ILP solutions of the following instances 2 (left) and 2a (right). The green line means two rolling pieces of stock coupled. We use $\alpha = 0.01$ in Eqs.~\eqref{eq::obj}–\eqref{eq::binary_req}.}
    \label{fig:train-diagrams-smaller}
\end{figure}

\begin{figure}[t!]
    \subfloat[]{\includegraphics[width=0.5\textwidth]{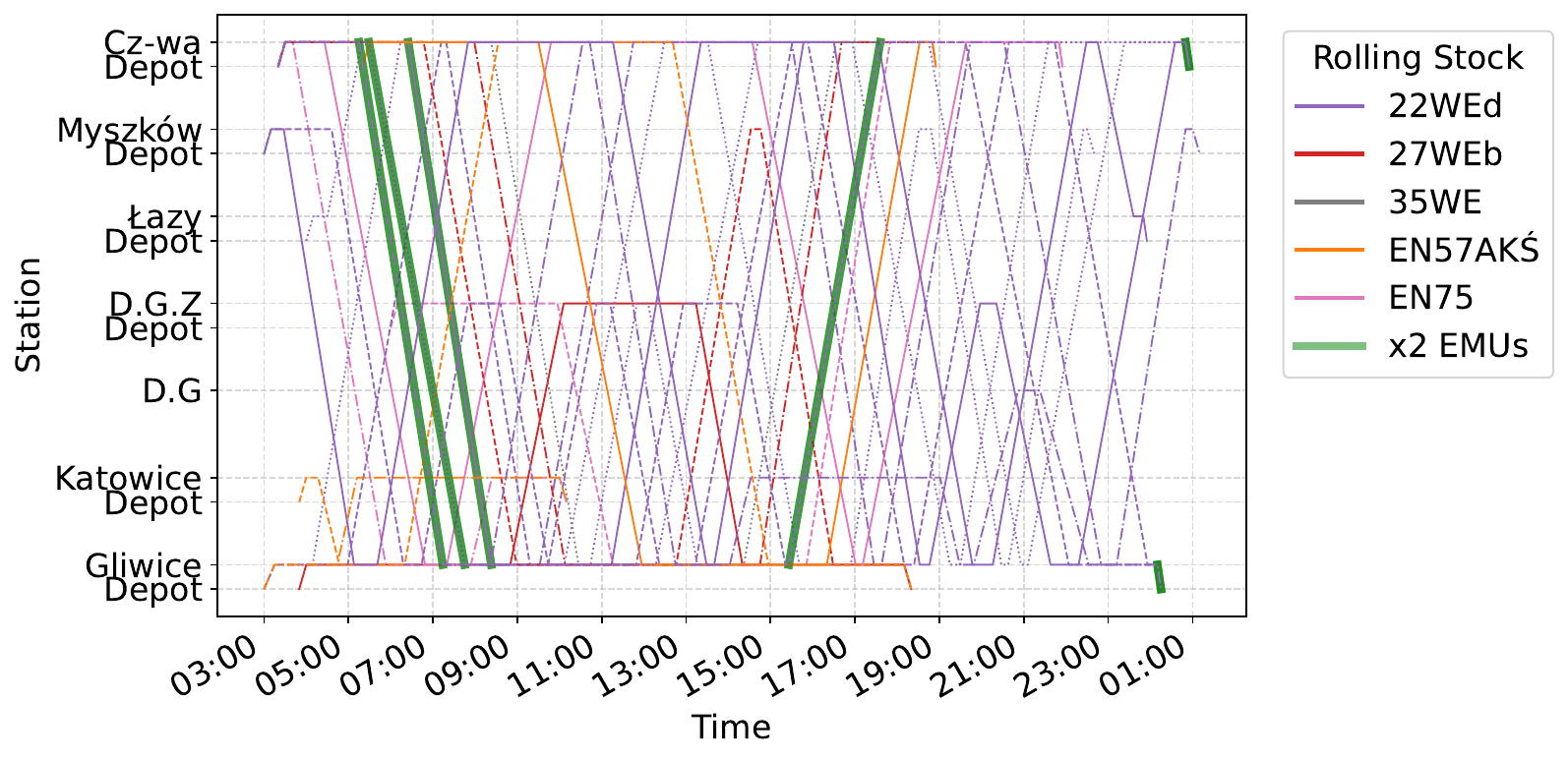}}%
    \subfloat[]{\includegraphics[width=0.5\textwidth]{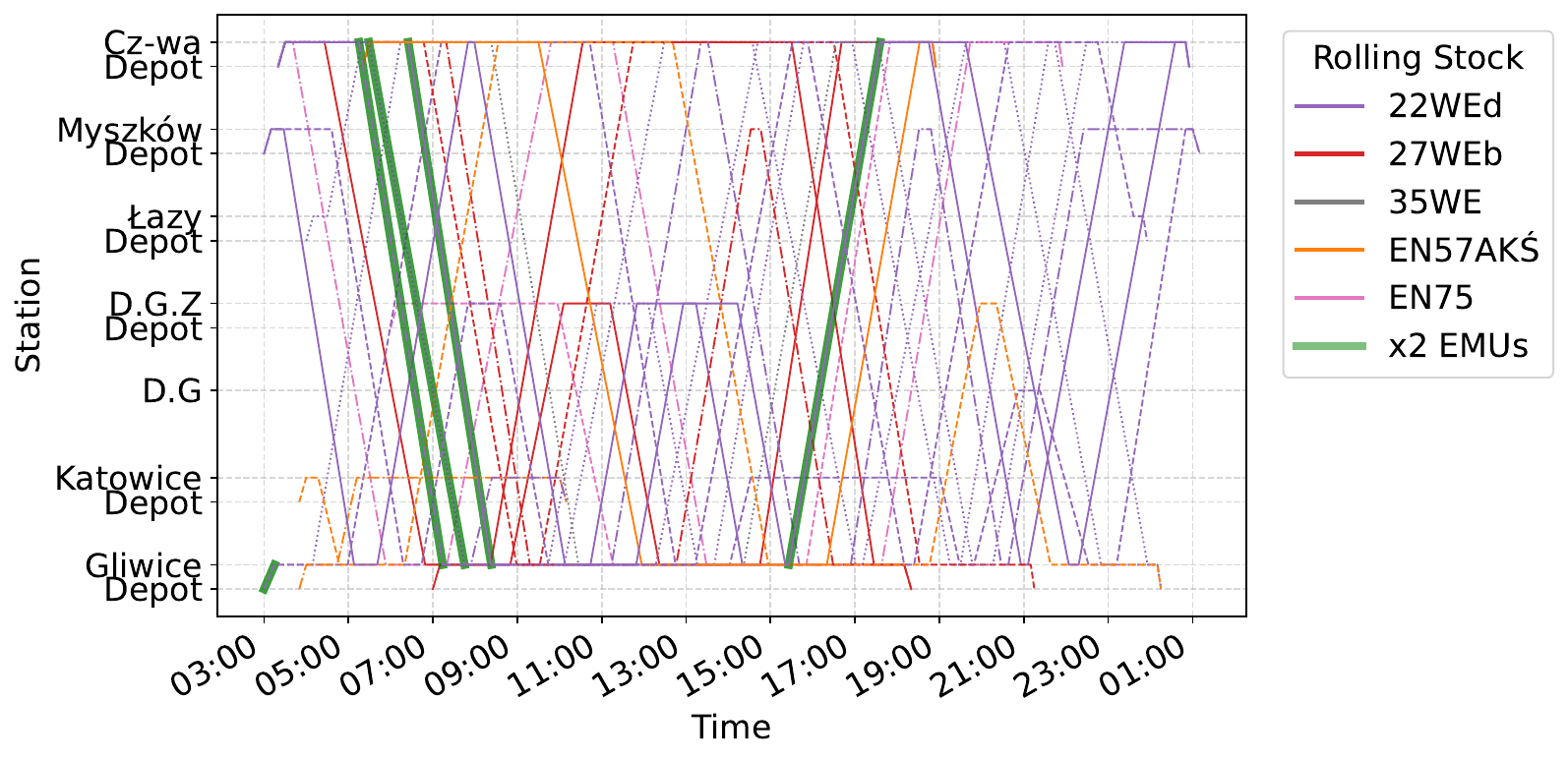}}
    \caption{Train diagrams of ILP solutions for instances 4 (left) and 4a (right). Different colors represent different types of rolling stock. The green line means two rolling pieces of stock coupled. We use $\alpha = 0.01$ in Eqs.~\eqref{eq::obj}–\eqref{eq::binary_req}.}
    \label{fig:train-diagrams}
\end{figure}

ILP solutions obtained using the SCIP solver \emph{version 9.2.1}~\cite{BolusaniEtal2024ZR} for these instances are presented in Table~\ref {tab::solutions}. Train diagrams for selected instances are presented in Fig.~\ref{fig:train-diagrams-smaller}, Fig.~\ref{fig:train-diagrams}. From a practical perspective, while respecting data confidentiality, we can reveal that Fig.~\ref{fig:train-diagrams} reproduces train runs involving multiple compositions, reasonably reflecting real-world practice and confirming the practical soundness of the model.

\begin{table}[t!]
\centering
\begin{tabular}{|l|l|l|l|l|l|l|}
	\hline
        & \multicolumn{2}{l|}{$\alpha = 0.01$} & \multicolumn{2}{l|}{$\alpha = 0.0001$}  &  \multicolumn{2}{l|}{$\alpha = 0.0$} \\ \hline
	inst. & \begin{tabular}{c} comp \\ time [s]  \end{tabular} &  objective &  \begin{tabular}{c} comp \\ time [s] \end{tabular} &  objective  &  \begin{tabular}{c} comp \\ time [s]  \end{tabular} &  objective  \\ \hline
	toy  &  $0$  &  $4.8$ &  $0$ &  $2.03$ &  $0$ & $2.0$   \\ \hline
	1 & $0$ & $110.66$ & $0$ & 2.10 &  $0$ & $1.0$   \\ \hline
        1a & $0$ & $119.71$ & $0$ & $3.18$ &  $0$ & $2.0$   \\ \hline
	2 & $0$ & $251.38$ &  $0$ & $5.48$   &  $0$ & $3.0$   \\ \hline
	2a & $0$ & $260.44$ &  $0$ & $6.56$ & $0$ & $4.0$  \\ \hline
        3 & $0.29$ & $2037.85$ &  $0.30$ & 3$6.22$  &  $0.91$ & $16.0$ \\ \hline
	4 & $2.17$ & $1817.56$ &  $3.23$ & $33.76$ & $5.44$ & $15.0$  \\ \hline
	4a & $1.88$ & $1843.56$ &  $1.89$ & $34.59$  & $4.63$ & $16.0$  \\ \hline
	5 &  $23.76$ & $2060.87$ & $26.23$ & $39.79$ &  $55.61$ & $19.0$  \\ \hline
	5a & $22.47$ & $2073.86$ & $22.61$ & $39.91$  &  $55.88$ & $19.0$  \\ \hline
        6 & $37.89$ & $2404.10$ & $31.15$ & $44.31$   & $31.50$ & $20.0$ \\ \hline
	6a & $33.14$ & $2416.53$ & $28.05$ & $44.55$  & $61.99$ & $20.0$  \\ \hline
	7 & $378.37$ & $4852.04$ &  $153.10$ & $92.18$  & $1137.89$ & $43.0$  \\ \hline
	7a & $278.28$ & $4853.11$ & $102.45$ & $92.26$ &  $911.95$ & $43.0$  \\ \hline
	8 & $250.28$ & $4759.22$ &  $698.92$ & $92.09$   & $2257.09$ & $43.0$  \\ \hline
	8a & $209.00$ & $4759.22$ &  $336.71$ & $92.17$  & $4852.04$ & $43.0$  \\ \hline
\end{tabular}
	\caption{ILP solutions of the SCIP solver \emph{version 9.2.1}. Computational time  rounded to $0.01$ s. Comparing cases 7, 7 a) with 8, 8 a) one can conclude that heuristic limitations of the $\Delta$ parameter reduce the computational time at a relatively small cost in the objective value. The computations were performed on 13th Gen Intel(R) Core(TM) i5-1340P @4 GHz, 12 cores, 16 threads.}
 	\label{tab::solutions}
\end{table}

The ILP formulation solved by SCIP scales well for practical daily railway instances. 
Even the largest considered instances (7--8a), containing approximately $3.8 \cdot 10^5$ binary variables, were solved within at most about one hour. 
For medium-size instances (3--6a), optimal solutions were typically obtained within seconds to tens of seconds.

The results also demonstrate the influence of the weighting coefficient $\alpha$ in the objective function. 
For $\alpha = 0$, the optimizer minimizes only the number of deployed EMUs, leading to solutions with the smallest fleet utilization but potentially higher operational costs. 
Increasing $\alpha$ shifts the optimization toward lower operating-cost circulations.

A comparison between instances 7 and 8 additionally illustrates the impact of the maximal transfer parameter $\Delta$. 
Increasing $\Delta$ from $300$ to $960$ significantly enlarges the feasible arc set and increases computational complexity, resulting in longer runtimes and larger ILP models.

\begin{figure}[t!]
    \centering
    \includegraphics[width=\linewidth]{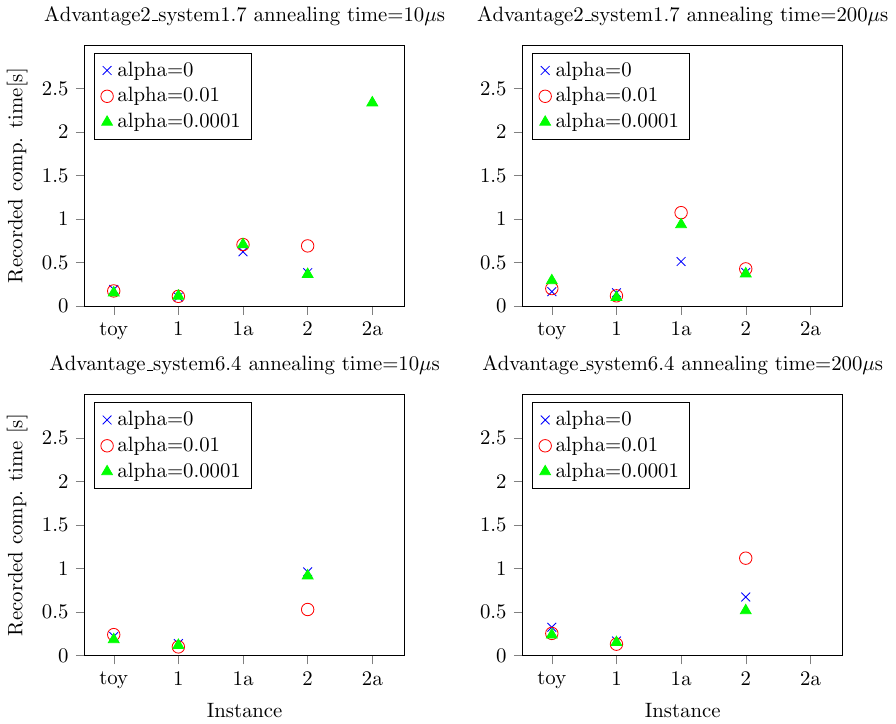}
    \caption{Results of the D-Wave experiments. We report the computation time ( minor embedding time, network/queue latency, QPU sampling time, embedding processing) for each instance together with the required number of runs to reach the optimal solution (we start with 100 runs on D-Wave and increase it gradually if the optimal solution is not found). In the 2a instance, we obtained, for one parameter setting, a solution that is optimal for the original problem but does not satisfy the constraints on the slack variables in the QUBO encoding. Instance~3 was too large to be embedded and thus could not be run on the D-Wave machine. If there is no record, we were unable to reach a feasible solution, within $5000$ runs of $10 \mu$s of annealing time or $2500$ runs of $200 \mu$s of annealing time. The D-Wave QPU was accessed through a 13th Gen Intel Core i7-13700HX with 24 threads and 64GB of RAM.}
    \label{fig::qa-solutions}
\end{figure}

The results of quantum annealing on the D-Wave machine are presented in Fig~\ref{fig::qa-solutions}. We have reached on the D-Wave machine the feasible and optimal solutions with objective value the same as in the ILP approach on the following instances: toy, $1$, $1$a, and $2$ for all settings. In one case, for instance, $2$a we have also obtained the feasible and optimal solution, but with broken constraints on the slack variables.

\subsection{Quantum annealing results}

Although the quantum annealing approach is currently applicable only to small instances and does not outperform the ILP approach in terms of computational time, there is clear progress in quantum annealing technology developments when comparing the older \texttt{Advantage\_system6.4} with the newer \texttt{Advantage2\_system1.7} \cite{dwave2025_adv2_whitepaper}. In the first case, only the simplest instances (toy, 1, 2) were solvable, while \texttt{Advantage2\_system1.7} allowed us to solve more complex ones, i.e., 1a and 2a. 

\begin{figure}[t!]
    \centering
    \subfloat[]{\includegraphics[width=0.5\textwidth]{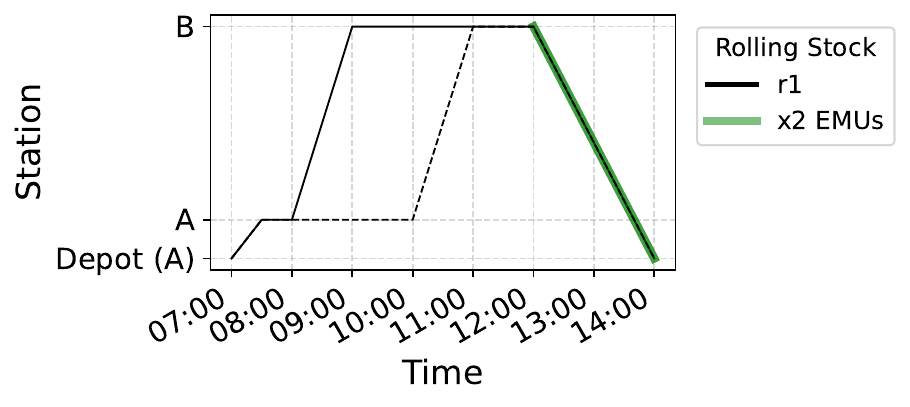}}%
    \subfloat[]{\includegraphics[width=0.5\textwidth]{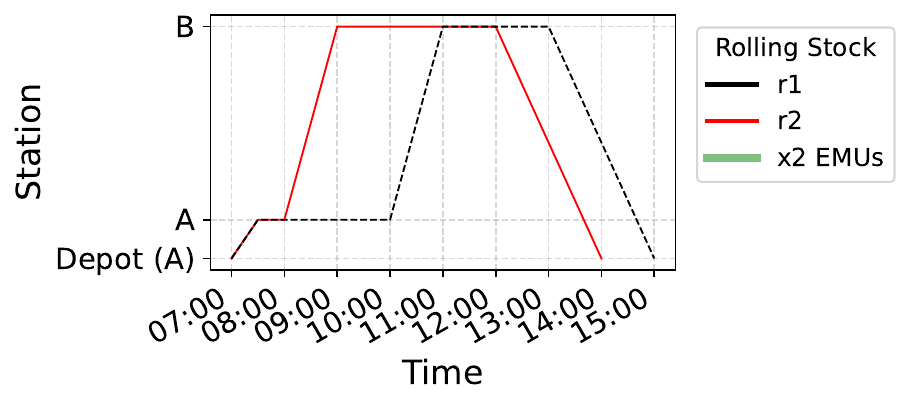}}
    \caption{Two D-Wave \texttt{Advantage2\_system1.7} solutions of the toy case: left, the ground state with objective $4.8$; right, an excited state with objective value $5.6$}
    \label{fig:toy_solutions}
\end{figure}

The \texttt{Advantage\_system6.4} uses the Pegasus topology (with $\sim$15-way connectivity) and has about 5,612 qubits, while the \texttt{Advantage2\_system1.7} uses the newer Zephyr topology with 20-way connectivity, enabling more compact embeddings. Advantage2 also has a higher energy scale, lower noise, and about twice the qubit coherence time, which together improve solution quality and speed \cite{dwave2025_press_adv2} \cite{dwave2025_adv2_whitepaper}. For various feasible solutions from quantum annealing, see the example on the toy model in Fig.~\ref{fig:toy_solutions}.

The experiments indicate that current quantum annealers are capable of solving only small instances of the considered railway circulation problem. 
Nevertheless, the results demonstrate clear progress between successive D-Wave hardware generations. 
While the older Pegasus-based system solved only the smallest instances, the newer Zephyr-based architecture additionally handled instances 1a and 2a.

The main practical limitation of the QUBO approach is not the optimization process itself, but the rapid growth in the number of quadratic terms required by the reformulation. 
This effect becomes dominant already for medium-sized railway instances.

\subsection{Quantum-inspired QUBO results}

Let us discuss the results obtained by the VeloxQ quantum-inspired solver~\cite{tuziemski2025veloxq} 
We observe better performance than D-Wave for small instances; however, large instances were not tractable with the QUBO approach because of the explosion of the number of terms. 
In fact, this is a general observation regarding the QUBO formulation: regardless of the QUBO solver applied, the main limitation on a bigger scale is the generation and storage of the QUBO model itself.
Even the generation of the model can take more time than the ILP solution. 

In those cases, however, when the QUBOs could be reasonably generated, VeloxQ gave promising results, outperforming DWave.
VeloxQ has two parameters: the number of time steps and the number of repetitions, which we have set to the default $4096$ and $5000$, respectively, resulting in a compute time of $0.2$ seconds per instance in the cloud.
Within this time, it has found the optimal solutions (as deduced from the exact ILP solutions) in all the cases which were tractable with the QUBO approach.
This illustrates that a good QUBO heuristic such as VeloxQ can be competitive even with ILP approaches, however, in the present problem the overhead in the problem generation poses a severe limitation: from instance $3$ on, the generation of the QUBOs was not reasonable.


\subsection{Analysis of Results}

The experiments demonstrate that the proposed ILP formulation is practically applicable to real-world regional railway circulation planning problems. 
The model successfully handles coupled EMU operations, depot balancing, passenger-capacity constraints, and driver limitations within realistic computational times.

The QUBO formulation currently remains limited to relatively small instances because of the rapid growth of quadratic interactions and embedding overheads. 
Nevertheless, the obtained results indicate that both quantum annealing and physics-inspired heuristics are capable of recovering high-quality feasible solutions for small and medium-size cases.

In practice, therefore, there is scope for applying these methods, as many operational disruptions stem from real traffic conditions. In such cases, the number of variables and conditions is significantly reduced compared to planning a schedule covering $400$ trips and several dozen vehicles. These are, however, situations that occur practically every day on the railway network (collisions with wildlife~\cite{gawlak2025statistical}, rolling stock failures, infrastructure malfunctions~\cite{koniorczyk2025solving}, etc.), which dispatchers currently have to handle on their own, making decisions in a very short amount of time based more on their experience than on the actual data they have.

From the practical perspective, the experiments confirm that restricting the transfer parameter $\Delta$ provides an effective heuristic reduction of the search space while preserving solution quality.

\section{Discussion}\label{sec::discussion}

We have examined daily rolling stock circulation planning for EMUs under operational constraints relevant to the Silesian Railway operator, including (i) limited, predefined coupling of identical units in pairs, and (ii) demand-driven capacity requirements for both seats and bicycle spaces. Using real timetable-based instances of increasing size, we have compared a classical ILP approach against QUBO-based solution paradigms—quantum annealing on D‑Wave and the quantum-inspired VeloxQ solver—to empirically characterize where QUBO methods are currently tractable for this application. 

For the tested daily instances, the classical ILP formulation can produce feasible circulation plans at practical computational cost across the full range of instance sizes considered. In particular, the ILP solves instances up to 404 trips and 11 EMU types (instances 7/7a/8/8a in Table~\ref{tab::instances}) with runtimes ranging from sub-second on small cases to tens of minutes on the largest cases depending on the objective weighting $\alpha$ (Table~\ref{tab::solutions}).

A practically relevant observation is that limiting the maximum transfer window $\Delta$—used as a heuristic to cap arc generation and thus reduce model density—can reduce solution time substantially with only a relatively small loss in objective value. This is visible when comparing cases 7/7a ($\Delta$ = 300) with 8/8a ($\Delta$ = 960) and is explicitly noted in Table~\ref{tab::solutions} caption. From an operational perspective, this supports a two-mode workflow: (i) a fast “restricted-$\Delta$” solve to obtain a high-quality plan quickly during disruptions, and (ii) an optional refinement step (larger $\Delta$) when additional computation time is available.

Across the tested instances, the dominant barrier for direct QUBO-based/Ising approaches is not the number of binary variables per se, but the quadratic expansion and connectivity/embedding overhead. In the larger instances, the number of QUBO variables remains close to the number of ILP variables (slack-variable overhead is limited), yet the number of QUBO terms grows rapidly, reaching $\sim 1.5 \times 10^8$ terms in instance 7/7a, and QUBO construction becomes infeasible for the largest $\Delta$ cases (8/8a) under the available 62~GB RAM. This practical bottleneck is consistent with the scaling discussion of the QUBO term count and with the observation that problem sparsification via tighter $\delta/\Delta$ windows is essential to keep QUBOs manageable.

On D‑Wave, quantum annealing is currently applicable only to the smallest instances considered. Instance 3 was already too large to embed, and for some settings (e.g., instance 2a), the sampler may return a solution that is optimal for the original objective but violates slack-variable constraints in the QUBO encoding—highlighting the sensitivity to penalty calibration and the need for systematic feasibility screening when using penalty-based formulations. At the same time, the comparison between hardware generations indicates tangible progress: the older Advantage\_system6.4 could solve only the simplest cases (toy, 1, 2), while Advantage2\_system1.7 enabled additional instances (1a and 2a for some parameter settings). For the VeloxQ solver, results are stronger than D‑Wave on small cases and were feasible and optimal for instances up to instance 3 under the default configuration; however, larger instances were still not tractable in this direct QUBO approach, and the remaining gap versus ILP remains substantial for practical deployment.

Regarding VeloxQ, a physics-inspired cloud-based heuristic solver: in those cases where QUBO generation was feasible within resource limits, the optimal solution and a number of near-optimal feasible alternatives were found within a short, predefined computational time, illustrating that QUBO solvers can be competitive --- but they must be applied to the right problem.

A key requirement from practice is not only to obtain one feasible plan, but to have multiple feasible alternatives available quickly so that dispatchers can account for non-modelled considerations and coordinate with infrastructure management. The toy example illustrates that the model naturally admits multiple feasible “excited” solutions beyond the optimum. This feature aligns well with the behavior of annealing-based approaches, which aim to sample from low-energy regions rather than output a single deterministic solution; in principle, this could be operationally valuable when a DSS needs to offer a small portfolio of near-optimal circulation options (e.g., different reserve usage levels, different coupling choices, or different depot return patterns) under time pressure. However, the experiments also show that obtaining a useful portfolio of feasible solutions from QUBO samplers requires careful constraint encoding (penalty design), feasibility checks, and likely post-processing—otherwise low-energy samples can still violate critical constraints (e.g., slack-variable feasibility).

Taken together, the results suggest a clear division of labor for near-term methods:
\begin{itemize}
    \item \textbf{Classical ILP} is currently the most reliable approach for end-to-end daily circulation planning at the tested scales (hundreds of trips, many depots, and many rolling-stock types).
    \item \textbf{QUBO-based Quantum and quantum-inspired} are presently most suitable as subproblem optimizers on carefully bounded instances (few rolling stock types, limited trip subsets, limited time windows), consistent with the limitations observed in embedding and QUBO construction.
\end{itemize}
This points toward hybrid architectures in which the full daily problem is solved classically, while QUBO-based optimization is applied selectively to localized, high-impact decisions. Concretely, promising decomposition directions include:
\begin{itemize}
    \item Local spatio-temporal neighborhoods around a disruption (e.g., the next 1–3 hours and a small corridor), where only a small set of trips and feasible turnarounds matter.
    \item Coupling/decoupling “hotspots” (stations and predefined coupling trips), where composition decisions create combinatorial difficulty but remain geographically and temporally localized.
    \item Type-restricted subproblems (a few EMU types relevant to the disrupted lines), which directly attack the observed $|R|$ (i.e. number of types of EMUs) sensitivity in QUBO term growth.
\end{itemize}
In such a hybrid approach, the quantum layer would not replace the ILP solver; rather, it would complement it by rapidly exploring alternative local rewirings (and producing multiple candidate solutions) while the classical layer maintains global feasibility and overall resource balance.

While daily (acyclic) planning appears tractable with classical solvers, extending to multi-day circulation introduces additional complexity that can quickly exceed straightforward daily formulations. \ref{appendix} sketches one potential direction inspired by column generation: start from daily schedules that begin/end in depots, then add variables that allow overnight transfers to join daily schedules into more efficient multi-day rotations under practical feasibility constraints. Developing this direction further—together with maintenance constraints, richer crew constraints, and disruption scenarios beyond demand surges—would provide a natural next step for both classical and hybrid solution approaches.

Finally, we note several limitations that motivate future work. The coupling policy is intentionally restricted (pairs of identical EMUs and only on predefined trips), which matches the partner’s operational practice but reduces flexibility. 
For QUBO-based methods, the study relies on penalty-based encodings and therefore inherits typical challenges of penalty selection, feasibility screening, and quadratic expansion; addressing these issues—e.g., by tighter formulations, structured decompositions, and systematic calibration—appears necessary before QUBO methods can be considered for larger real-world circulation instances.

\section*{Acknowledgements}
The authors acknowledge the cooperation with the railway operator \href{https://www.kolejeslaskie.pl/en/}{Koleje Śląskie sp. z o.o}, in particular concerning the problem statement, data acquisition, and discussion on results.
The authors acknowledge the \href{https://www.fz-juelich.de/ias/jsc}{J\"ulich Supercomputing Centre} for providing computing time on the D-Wave Advantage™ System JUPSI through the J\"ulich UNified Infrastructure for Quantum computing (JUNIQ).

Z.M. acknowledges funding from the Ministry of Economic Affairs, Labour and Tourism Baden-Württemberg in the frame of the Competence Center Quantum Computing Baden-Württemberg (project ``KQCBW25'')

K.D. acknowledges: Scientific work co-financed from the state budget under the program of the Minister of Education
and Science, Poland (pl. Polska) under the name "Science for Society II" project number NdS-II/SP/0336/2024/01 funding amount 1000000 PLN  
total value of the project 1000000 PLN  \raisebox{-4pt}{\includegraphics[width = 0.15\textwidth]{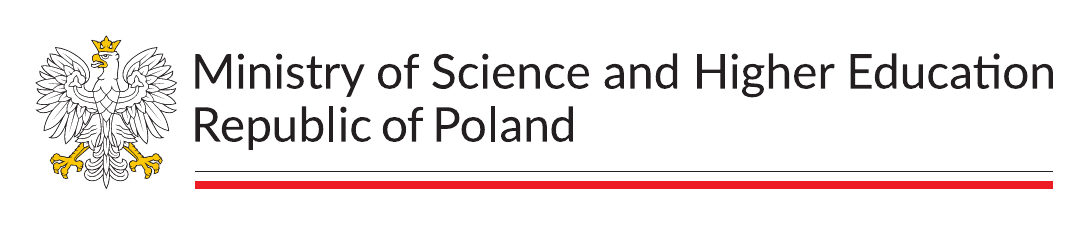}}

M.K. acknowledges the support of the Research Excellence Programme of the National Research, Development, and Innovation Office (NKFIH) of Hungary (Grant No. KKP133827). 

We appreciate the assistance of Peter M\'arton - University of \v{Z}ilina,
Department of Mathematical Methods and Operations Research, Faculty of Management Science and Informatics - on discussing the railway model and feedback on the manuscript. We would like to thank Bartłomiej Gardas and Łukasz Pawela for supplying access to VeloxQ solver~\cite{tuziemski2025veloxq}.

\section*{Declaration of generative AI and AI-assisted technologies in the manuscript preparation process}


During the preparation of this work, the authors used ChatGPT, developed by OpenAI, and Claude, developed by Anthropic, to support manuscript preparation, including improving the structure of selected sections, enhancing the clarity and flow of English-language text, and making descriptions more readable and coherent. After using these tools, the authors reviewed, edited, and verified the content as needed and take full responsibility for the content of the published article.

\appendix 
\section{Multi-day extension of the model}\label{appendix}

\paragraph{Potential multi-day EMUs circulation plan} Although our model is primarily designed for daily plans, it can be modified for the multi-day circulation plans. Here, our partner limits himself to a rather short few-day schedule with a limit of say $5$ days. To solve such a problem, inspired by the column generation approach~\cite{janacek2017optimization}, in terms of adding new variables to join one-day schedules. 
\begin{itemize}
\item We start with the sequence of one-day schedules where all trains have one-day trips starting and ending in dedicated depots.
\item Then we create new variables that allow for the transfer of the rolling stock overnight in some chosen locations, joining one-day schedules into the multi-day ones that are more effective (e.g., in terms of used rolling stock)
\item to select the new variables, we use the subset of operation points satisfying the partner's requirements and the technical feasibility of staying of rolling stock overnight.
\end{itemize}


\end{document}